\documentclass[twocolumn, longauth]{aa}
\usepackage{graphicx}
\usepackage[varg]{txfonts}
\usepackage[colorlinks = true,
            linkcolor = blue,
            urlcolor  = blue,
            citecolor = blue,
            anchorcolor = blue]{hyperref}
\usepackage{euclid}
\usepackage{bm}
\usepackage{soul}
\usepackage[dvipsnames]{xcolor}
\usepackage{verbatim}
\def\be{\begin{equation}}
\def\ee{\end{equation}}
\def\ba{\begin{eqnarray}}
\def\ea{\end{eqnarray}}
\def\eqi{\begin{equation}}
\def\eqf{\end{equation}}
\def\eqia{\begin{eqnarray}}
\def\eqfa{\end{eqnarray}}
\def\lcdm{$\Lambda$CDM }

\usepackage{ulem}
\usepackage{multirow}
\usepackage{booktabs}
\usepackage{relsize}
\usepackage{nameref}
\usepackage{array}
\usepackage{float}
\usepackage{stfloats}

\usepackage[nameinlink,noabbrev]{cleveref}
\Crefname{equation}{Eq.}{Eqs.}
\Crefname{eqnarray}{Eq.}{Eqs.}
\Crefname{section}{Sect.}{Sects.}
\Crefname{figure}{Fig.}{Figs.}
\crefname{equation}{Equation}{Equations}
\crefname{section}{Section}{Sections}
\crefname{figure}{Figure}{Figures}

\usepackage{enumitem}

\SetLabelAlign{Center}{\hfil#1\hfil}
\SetLabelAlign{CenterWithParen}{\hfil(\makebox[1.0em]{#1})\hfil}

\DeclareRobustCommand{\rchi}{{\mathpalette\irchi\relax}}
\newcommand{\irchi}[2]{\raisebox{\depth}{$#1\chi$}}
\newcolumntype{C}[1]{>{\centering\arraybackslash}p{#1}}

\begin{document}

\title{\Euclid: Cosmological forecasts from the void size function\thanks{This paper is published on behalf of the Euclid Consortium.}}

\titlerunning{Forecasts on void size function constraints}

\newcommand{\orcid}[1]{} 
\author{S.~Contarini\orcid{0000-0002-9843-723X}$^{1,2,3}$\thanks{\email{sofia.contarini3@unibo.it}}, G.~Verza\orcid{0000-0002-1886-8348}$^{4,5}$\thanks{\email{giovanni.verza@pd.infn.it}}, A.~Pisani\orcid{0000-0002-6146-4437}$^{6,7,8}$, N.~Hamaus\orcid{0000-0002-0876-2101}$^{9}$, M.~Sahl\'en\orcid{0000-0003-0973-4804}$^{10,11}$, C.~Carbone$^{12}$, S.~Dusini\orcid{0000-0002-1128-0664}$^{4}$, F.~Marulli\orcid{0000-0002-8850-0303}$^{1,2,3}$, L.~Moscardini\orcid{0000-0002-3473-6716}$^{1,2,3}$, A.~Renzi\orcid{0000-0001-9856-1970}$^{4,5}$
% ...
}

\institute{$^{1}$ Dipartimento di Fisica e Astronomia "Augusto Righi" - Alma Mater Studiorum Universit\`{a} di Bologna, via Piero Gobetti 93/2, I-40129 Bologna, Italy\\
$^{2}$ INFN-Sezione di Bologna, Viale Berti Pichat 6/2, I-40127 Bologna, Italy\\
$^{3}$ INAF-Osservatorio di Astrofisica e Scienza dello Spazio di Bologna, Via Piero Gobetti 93/3, I-40129 Bologna, Italy\\
$^{4}$ INFN-Padova, Via Marzolo 8, I-35131 Padova, Italy\\
$^{5}$ Dipartimento di Fisica e Astronomia "G.Galilei", Universit\'a di Padova, Via Marzolo 8, I-35131 Padova, Italy\\
$^{6}$ Department of Astrophysical Sciences, Peyton Hall, Princeton University, Princeton, NJ 08544, USA\\
$^{7}$ Center for Computational Astrophysics, Flatiron Institute, 162 5th Avenue, 10010, New York, NY, USA\\
$^{8}$ The Cooper Union for the Advancement of Science and Art, 41 Cooper Square, New York, NY 10003, USA\\
$^{9}$ Universit\"ats-Sternwarte M\"unchen, Fakult\"at f\"ur Physik, Ludwig-Maximilians-Universit\"at M\"unchen, Scheinerstrasse 1, 81679 M\"unchen, Germany\\
$^{10}$ Theoretical astrophysics, Department of Physics and Astronomy, Uppsala University, Box 515, SE-751 20 Uppsala, Sweden\\
$^{11}$ Swedish Collegium for Advanced Study, Thunbergsv\"{a}gen 2, SE-752 38 Uppsala, Sweden\\
$^{12}$ INAF-IASF Milano, Via Alfonso Corti 12, I-20133 Milano, Italy\\
$^{13}$ University of Lyon, UCB Lyon 1, CNRS/IN2P3, IUF, IP2I Lyon, France\\
$^{14}$ Aix-Marseille Univ, CNRS/IN2P3, CPPM, Marseille, France\\
$^{15}$ Instituto de Astrof\'isica de Canarias (IAC); Departamento de Astrof\'isica, Universidad de La Laguna (ULL), E-38200, La Laguna, Tenerife, Spain\\
$^{16}$ Departamento de Astrof\'{i}sica, Universidad de La Laguna, E-38206, La Laguna, Tenerife, Spain\\
$^{17}$ Institut d'Astrophysique de Paris, UMR 7095, CNRS, and Sorbonne Universit\'e, 98 bis boulevard Arago, 75014 Paris, France\\
$^{18}$ Centre for Astrophysics, University of Waterloo, Waterloo, Ontario N2L 3G1, Canada\\
$^{19}$ Department of Physics and Astronomy, University of Waterloo, Waterloo, Ontario N2L 3G1, Canada\\
$^{20}$ Institute of Cosmology and Gravitation, University of Portsmouth, Portsmouth PO1 3FX, UK\\
$^{21}$ SISSA, International School for Advanced Studies, Via Bonomea 265, I-34136 Trieste TS, Italy\\
$^{22}$ IFPU, Institute for Fundamental Physics of the Universe, via Beirut 2, 34151 Trieste, Italy\\
$^{23}$ Univ Lyon, Univ Claude Bernard Lyon 1, CNRS/IN2P3, IP2I Lyon, UMR 5822, F-69622, Villeurbanne, France\\
$^{24}$ Institut de Recherche en Astrophysique et Plan\'etologie (IRAP), Universit\'e de Toulouse, CNRS, UPS, CNES, 14 Av. Edouard Belin, F-31400 Toulouse, France\\
$^{25}$ Universit\'e St Joseph; Faculty of Sciences, Beirut, Lebanon\\
$^{26}$ Department of Mathematics and Physics, Roma Tre University, Via della Vasca Navale 84, I-00146 Rome, Italy\\
$^{27}$ Institut d'Astrophysique de Paris, 98bis Boulevard Arago, F-75014, Paris, France\\
$^{28}$ Dipartimento di Fisica e Astronomia, Universit\'a di Bologna, Via Gobetti 93/2, I-40129 Bologna, Italy\\
$^{29}$ INAF-Osservatorio Astrofisico di Torino, Via Osservatorio 20, I-10025 Pino Torinese (TO), Italy\\
$^{30}$ Dipartimento di Fisica, Universit\'a degli studi di Genova, and INFN-Sezione di Genova, via Dodecaneso 33, I-16146, Genova, Italy\\
$^{31}$ INFN-Sezione di Roma Tre, Via della Vasca Navale 84, I-00146, Roma, Italy\\
$^{32}$ INAF-Osservatorio Astronomico di Capodimonte, Via Moiariello 16, I-80131 Napoli, Italy\\
$^{33}$ Instituto de Astrof\'isica e Ci\^encias do Espa\c{c}o, Universidade do Porto, CAUP, Rua das Estrelas, PT4150-762 Porto, Portugal\\
$^{34}$ Dipartimento di Fisica, Universit\'a degli Studi di Torino, Via P. Giuria 1, I-10125 Torino, Italy\\
$^{35}$ INFN-Sezione di Torino, Via P. Giuria 1, I-10125 Torino, Italy\\
$^{36}$ Port d'Informaci\'{o} Cient\'{i}fica, Campus UAB, C. Albareda s/n, 08193 Bellaterra (Barcelona), Spain\\
$^{37}$ Institut de F\'{i}sica d'Altes Energies (IFAE), The Barcelona Institute of Science and Technology, Campus UAB, 08193 Bellaterra (Barcelona), Spain\\
$^{38}$ INAF-Osservatorio Astronomico di Roma, Via Frascati 33, I-00078 Monteporzio Catone, Italy\\
$^{39}$ INFN section of Naples, Via Cinthia 6, I-80126, Napoli, Italy\\
$^{40}$ Department of Physics "E. Pancini", University Federico II, Via Cinthia 6, I-80126, Napoli, Italy\\
$^{41}$ Institut national de physique nucl\'eaire et de physique des particules, 3 rue Michel-Ange, 75794 Paris C\'edex 16, France\\
$^{42}$ Institute for Astronomy, University of Edinburgh, Royal Observatory, Blackford Hill, Edinburgh EH9 3HJ, UK\\
$^{43}$ Jodrell Bank Centre for Astrophysics, Department of Physics and Astronomy, University of Manchester, Oxford Road, Manchester M13 9PL, UK\\
$^{44}$ ESAC/ESA, Camino Bajo del Castillo, s/n., Urb. Villafranca del Castillo, 28692 Villanueva de la Ca\~nada, Madrid, Spain\\
$^{45}$ European Space Agency/ESRIN, Largo Galileo Galilei 1, 00044 Frascati, Roma, Italy\\
$^{46}$ Institute of Physics, Laboratory of Astrophysics, Ecole Polytechnique F\'{e}d\'{e}rale de Lausanne (EPFL), Observatoire de Sauverny, 1290 Versoix, Switzerland\\
$^{47}$ Mullard Space Science Laboratory, University College London, Holmbury St Mary, Dorking, Surrey RH5 6NT, UK\\
$^{48}$ Departamento de F\'isica, Faculdade de Ci\^encias, Universidade de Lisboa, Edif\'icio C8, Campo Grande, PT1749-016 Lisboa, Portugal\\
$^{49}$ Instituto de Astrof\'isica e Ci\^encias do Espa\c{c}o, Faculdade de Ci\^encias, Universidade de Lisboa, Campo Grande, PT-1749-016 Lisboa, Portugal\\
$^{50}$ Department of Astronomy, University of Geneva, ch. d\'Ecogia 16, CH-1290 Versoix, Switzerland\\
$^{51}$ Department of Physics, Oxford University, Keble Road, Oxford OX1 3RH, UK\\
$^{52}$ AIM, CEA, CNRS, Universit\'{e} Paris-Saclay, Universit\'{e} de Paris, F-91191 Gif-sur-Yvette, France\\
$^{53}$ Institut d'Estudis Espacials de Catalunya (IEEC), Carrer Gran Capit\'a 2-4, 08034 Barcelona, Spain\\
$^{54}$ Institute of Space Sciences (ICE, CSIC), Campus UAB, Carrer de Can Magrans, s/n, 08193 Barcelona, Spain\\
$^{55}$ INAF-Osservatorio Astronomico di Trieste, Via G. B. Tiepolo 11, I-34143 Trieste, Italy\\
$^{56}$ Istituto Nazionale di Astrofisica (INAF) - Osservatorio di Astrofisica e Scienza dello Spazio (OAS), Via Gobetti 93/3, I-40127 Bologna, Italy\\
$^{57}$ Istituto Nazionale di Fisica Nucleare, Sezione di Bologna, Via Irnerio 46, I-40126 Bologna, Italy\\
$^{58}$ INAF-Osservatorio Astronomico di Padova, Via dell'Osservatorio 5, I-35122 Padova, Italy\\
$^{59}$ Max Planck Institute for Extraterrestrial Physics, Giessenbachstr. 1, D-85748 Garching, Germany\\
$^{60}$ Dipartimento di Fisica "Aldo Pontremoli", Universit\'a degli Studi di Milano, Via Celoria 16, I-20133 Milano, Italy\\
$^{61}$ INAF-Osservatorio Astronomico di Brera, Via Brera 28, I-20122 Milano, Italy\\
$^{62}$ INFN-Sezione di Milano, Via Celoria 16, I-20133 Milano, Italy\\
$^{63}$ Institute of Theoretical Astrophysics, University of Oslo, P.O. Box 1029 Blindern, N-0315 Oslo, Norway\\
$^{64}$ Jet Propulsion Laboratory, California Institute of Technology, 4800 Oak Grove Drive, Pasadena, CA, 91109, USA\\
$^{65}$ von Hoerner \& Sulger GmbH, Schlo{\ss}Platz 8, D-68723 Schwetzingen, Germany\\
$^{66}$ Max-Planck-Institut f\"ur Astronomie, K\"onigstuhl 17, D-69117 Heidelberg, Germany\\
$^{67}$ Universit\'e Paris-Saclay, Universit\'e Paris Cit\'e, CEA, CNRS, Astrophysique, Instrumentation et Mod\'elisation Paris-Saclay, 91191 Gif-sur-Yvette, France\\
$^{68}$ Universit\'e de Gen\`eve, D\'epartement de Physique Th\'eorique and Centre for Astroparticle Physics, 24 quai Ernest-Ansermet, CH-1211 Gen\`eve 4, Switzerland\\
$^{69}$ Department of Physics and Helsinki Institute of Physics, Gustaf H\"allstr\"omin katu 2, 00014 University of Helsinki, Finland\\
$^{70}$ European Space Agency/ESTEC, Keplerlaan 1, 2201 AZ Noordwijk, The Netherlands\\
$^{71}$ NOVA optical infrared instrumentation group at ASTRON, Oude Hoogeveensedijk 4, 7991PD, Dwingeloo, The Netherlands\\
$^{72}$ Argelander-Institut f\"ur Astronomie, Universit\"at Bonn, Auf dem H\"ugel 71, 53121 Bonn, Germany\\
$^{73}$ Department of Physics, Institute for Computational Cosmology, Durham University, South Road, DH1 3LE, UK\\
$^{74}$ University of Applied Sciences and Arts of Northwestern Switzerland, School of Engineering, 5210 Windisch, Switzerland\\
$^{75}$ Department of Physics and Astronomy, University of Aarhus, Ny Munkegade 120, DK-8000 Aarhus C, Denmark\\
$^{76}$ Perimeter Institute for Theoretical Physics, Waterloo, Ontario N2L 2Y5, Canada\\
$^{77}$ Space Science Data Center, Italian Space Agency, via del Politecnico snc, 00133 Roma, Italy\\
$^{78}$ Centre National d'Etudes Spatiales, Toulouse, France\\
$^{79}$ Institute of Space Science, Bucharest, Ro-077125, Romania\\
$^{80}$ Aix-Marseille Univ, CNRS, CNES, LAM, Marseille, France\\
$^{81}$ Centro de Investigaciones Energ\'eticas, Medioambientales y Tecnol\'ogicas (CIEMAT), Avenida Complutense 40, 28040 Madrid, Spain\\
$^{82}$ Instituto de Astrof\'isica e Ci\^encias do Espa\c{c}o, Faculdade de Ci\^encias, Universidade de Lisboa, Tapada da Ajuda, PT-1349-018 Lisboa, Portugal\\
$^{83}$ Universidad Polit\'ecnica de Cartagena, Departamento de Electr\'onica y Tecnolog\'ia de Computadoras, 30202 Cartagena, Spain\\
$^{84}$ Kapteyn Astronomical Institute, University of Groningen, PO Box 800, 9700 AV Groningen, The Netherlands\\
$^{85}$ Infrared Processing and Analysis Center, California Institute of Technology, Pasadena, CA 91125, USA\\
$^{86}$  Universit\'e Paris Cit\'e, CNRS, Astroparticule et Cosmologie, F-75013 Paris, France}

\authorrunning{S. Contarini et al.}

\abstract{The \Euclid mission -- with its spectroscopic galaxy survey covering a sky area over $15\,000 \ \mathrm{deg}^2$ in the redshift range $0.9<z<1.8$ -- will provide a sample of tens of thousands of cosmic voids. This paper thoroughly explores for the first time the constraining power of the void size function on the properties of dark energy (DE) from a survey mock catalogue, the official \Euclid Flagship simulation. We identified voids in the Flagship light-cone, which closely matches the features of the upcoming \Euclid spectroscopic data set. We modelled the void size function considering a state-of-the art methodology: we relied on the volume-conserving (Vdn) model, a modification of the popular Sheth \& van de Weygaert model for void number counts, extended by means of a linear function of the large-scale galaxy bias. We found an excellent agreement between model predictions and measured mock void number counts. We computed updated forecasts for the \Euclid mission on DE from the void size function and provided reliable void number estimates to serve as a basis for further forecasts of cosmological applications using voids. We analysed two different cosmological models for DE: the first described by a constant DE equation of state parameter, $w$, and the second by a dynamic equation of state with coefficients $w_0$ and $w_a$. 
We forecast $1\sigma$ errors on $w$ lower than $10\%$ and we estimated an expected figure of merit (FoM) for the dynamical DE scenario $\mathrm{FoM}_{w_0,w_a} = 17$ when considering only the neutrino mass as additional free parameter of the model. The analysis is based on conservative assumptions to ensure full robustness, and is a pathfinder for future enhancements of the technique. Our results showcase the impressive constraining power of the void size function from the \Euclid spectroscopic sample, both as a stand-alone probe, and to be combined with other \Euclid cosmological probes. \\
}

\maketitle

\section{Introduction} \label{sec:intro}

Cosmic voids are vast under-dense regions filling most of the volume of the present-day Universe. 
With sizes up to hundreds of megaparsec \citep{Gregory_1978,Tikhonov2006, Thompson2011, Szapudi2015} they are the largest observable structures in the cosmic web \citep{zeldovich1982,bond1996} -- the pattern arising in the galaxy distribution. 
Voids constitute a unique cosmological probe: their interiors, spanning a large range of scales and featuring low matter density, make them particularly suited to study dark energy (DE) and modified gravity \citep{lee2009,biswas2010,liMG2010,clampittMG2013,spolyarMG2013,cai2015,pisani2015,pollina2015,zivickMG2015,achitouv2016,sahlen2016,falck2018,sahlen2018,paillas2019,perico2019,verza2019,Contarini2020}, as well as massive neutrinos \citep{massara2015,banerjee2016,kreisch2018,sahlen2019,schuster2019, Kreisch2021}, primordial non-Gaussianity \citep{chan2019}, and physics beyond the standard model \citep{peebles2001,reed2015,yang2015,baldi2016}. 
Cosmic voids are becoming an effective and competitive new probe of cosmology thanks to the advent of current and upcoming sky surveys such as 6dF Galaxy Survey \citep[6dFGS,][]{6DF_jones2009}, VIMOS Public Extragalactic Redshift Survey \citep[VIPERS,][]{VIPERS_Guzzo2014}, SDSS-III's Baryon Oscillation Spectroscopic Survey \citep[BOSS,][]{BOSS_alam2017} and Extended Baryon Oscillation Spectroscopic Survey \citep[eBOSS,][]{eBOSS_dawson2016} from the Sloan Digital Sky Survey \citep[SDSS,][]{SDSS_blanton2017}, Dark Energy Survey \citep[DES,][]{DES2016}, Dark Energy Spectroscopic Instrument \citep[DESI,][]{DESI2016}, Prime Focus Spectrograph \citep[PFS,][]{PFS_2016}, the Roman Space Telescope \citep{spergel_2015_WFIRST}, Spectro-Photometer for the History of the Universe and Ices Explorer \citep[SPHEREx,][]{SPHEREx_2018}, and Large Synoptic Survey Telescope \citep[LSST,][]{LSST_ivezic2019}. 
Studying voids requires redshift surveys of very large volume, deep enough in the red band to measure a huge number of redshifts also for low-mass galaxies, and to map in detail significant contiguous fractions of the observable Universe. 
The \Euclid survey, expected to sample the sky over $15\,000 \ \mathrm{deg}^2$, will provide a unique opportunity to capitalise on cosmic voids, to leverage on measurements of the galaxy distribution at large scales and to improve our knowledge on cosmology and fundamental physics. Voids hold the keys to 
shed light on some of today’s open problems in cosmology \citep[][and references therein]{pisani2019}.

Cosmic voids from recent galaxy surveys have been used in a wide range of cosmological applications. They are sensitive to geometric effects, such as the Alcock--Paczy{\'n}ski effect \citep{AP1979,lavaux2012,sutter2012,sutter2014,hamaus2016,mao20217} and baryonic acoustic oscillations \citep{Kitaura2016, liang2016,chan2020,ForeroSanchez2021,Khoraminezhad2022}, as well as redshift-space distortions \citep[RSD,][]{paz2013,hamaus2014b,hamaus2015,cai2016,chuang2017,achitouv2017,hawken2017,hamaus2017,correa2019,achitouv2019,nadathur2019,nadathur2019b,hawken2020,hamaus2020,nadathur2020,Correa2021,aubert2020}, weak lensing \citep{melchior2014,clampitt2015,chantavat2016,gruen2016,chantavat2017,cai2017,sanchez2017,baker2018,brouwer2018,fang_DES2019,vielzeuf2019,Davis2021,Bonici2022}, the integrated Sachs–Wolfe effect and cross-correlation with the cosmic microwave background \citep[CMB,][]{Baccigalupi1999,granett2008,papai2010,cai2010,nadathur2012,flender2013,ilic2013,cai2014,cai2014b,nadathur2016,cai2017,kovacs2017,kovacs2018,kovacs2019,dong2020,hang_2021,kovacs2021,Kovacs2021b}. See e.g. \citet{pisani2019} and \citet{Moresco2022} for a review on cosmic void applications for cosmology.

In this paper we consider the void size function, which describes the number density of voids as a function of their size. 
Over the last two decades, studies of the hierarchical evolution of the void population in the excursion-set framework have allowed the construction of a theoretical void size function model built from first principles, 
the so-called Sheth \& van de Weygaert model \citep{SVdW2004}, later extended by \citet{jennings2013}. The void size function and its link to voids detected in galaxy surveys have been explored in depth with cosmological simulations \citep{furlanetto2006,platen2007,paranjape2012,jennings2013,
pisani2015,ronconi2019,contarini2019,verza2019,Contarini2020} and recently this statistic has proved to be a promising tool to constrain cosmology \citep{pisani2015,sahlen2019,contarini2019,verza2019,Kreisch2021}. 
The void size function has already been measured in surveys \citep[see e.g.][]{nadathur2016b,mao2016,aubert2020,hamaus2020}, and used for extreme-value statistics cosmology constraints \citep{sahlen2016}, but the void size function as a stand-alone probe has not yet been used to derive cosmological constraints.

In this work we focus on the power of the void size function from the \Euclid survey to constrain cosmological parameters. This study relies on the largest \Euclid-like light-cone, the \textit{Flagship} simulation \citep{Potter2017}.
The paper belongs to a series of companion papers investigating the scientific return that can be expected from voids observed by the \Euclid mission.
It aims at measuring and theoretically modelling the void size function from the Flagship simulation, providing a state-of-the-art forecast for void numbers to be expected from the \Euclid survey. Our model allows us to estimate the constraining power of the void size function on the DE equation of state while also varying the total matter density of the Universe and the total mass of neutrinos.
This analysis is focused on voids found in the spectroscopic galaxy distribution, for which the identification of voids is particularly accurate and reliable. We note that we leave for future work the measurement of the void size function in the photometric galaxy distribution from \Euclid, for which the data treatment greatly differs from the spectroscopic one (see e.g. \citealp{pollina2019} and \citealp[]{Bonici2022}).

The paper is organised as follows: in \Cref{sec:sim_voidFinding} we introduce the Flagship simulation, and describe the void finder and the cleaning algorithm used to obtain the void catalogue; in \Cref{sec:theory_&_methods} we present the theoretical model of the void size function (\Cref{subsec:theory_Vdn}), describe how to self-consistently align the measured void catalogue with the theoretical description (\Cref{subsec:methodology}), discuss the Bayesian statistical analysis used to perform the cosmological forecasts (\Cref{subsec:forecasts_theory}), and finally introduce the cosmological models considered in this work (\Cref{subsec:cosmological_models}). In \Cref{sec:results} we fit the theoretical model to the measured void size function in the Flagship simulation (\Cref{subsec:Vdn_analysis}) to obtain constraints on the DE equation of state and the remaining considered cosmological parameters, for different adopted approaches (\Cref{subsec:cosmological_forecats}); we conclude giving a discussion and a summary of our results in \Cref{sec:conclusions}.

\section{Galaxy and void catalogues}\label{sec:sim_voidFinding}

We now introduce the main tools for our work: the simulation and the void catalogues. This section also includes a brief description of the void finder and of the void catalogue preparation.

\subsection{Flagship simulation}\label{subsec:Flagship}
In this work we employed the \Euclid Flagship mock galaxy catalogue\footnote{Version $\mathrm{1.8.4}$.} (\textcolor{blue}{Castander et al., in prep.}). This catalogue was created running a simulation of two trillion dark matter particles in a periodic box of $L=3780 \ \hMpc$ per side \citep{Potter2017}, with a flat $\Lambda$-cold dark matter ($\Lambda$CDM) cosmology characterised by the parameters $\Omega_{\rm m}=0.319$, $\Omega_{\rm b}=0.049$, $\Omega_{\rm de}=0.681$, $\sigma_8=0.83$, $n_{\rm s}=0.96$ and $h=0.67$, as obtained by \textit{Planck} in 2018 \citep{Planck2018}. The simulation box was converted into a light-cone and the dark matter haloes have been identified using the \texttt{Rockstar} halo finder \citep{Behroozi2013}. These haloes were populated with central and satellite galaxies using a halo occupation distribution (HOD) method, to reproduce all the observables relevant for \textit{Euclid}'s main cosmological probes. Specifically, the HOD algorithm was calibrated exploiting several local observational constraints, using for instance the local luminosity function for the faintest galaxies \citep{Blanton2003, Blanton2005} and the galaxy clustering as a function of luminosity and colour \citep{Zehavi2011}. This galaxy sample is composed of more than two billion objects and presents a cut at magnitude $H<26$ or on the H$\alpha$ flux $f_{\mathrm{H}\alpha} > 2 \times 10^{-16} \ \mathrm{ergs} \ \mathrm{s}^{-1} \ \mathrm{cm}^{-2}$, which mimics the observation range expected for \Euclid. To match the completeness and the spectroscopic performance expected for the \Euclid survey, we uniformly downsampled the galaxy catalogue to consider only 60\% of the galaxies originally included in it. Furthermore we assumed our galaxy sample to have a purity of $100 \%$ and associated a Gaussian error of $\sigma_z = 0.001$ to the redshift of each galaxy \citep{EC2020}.
The full catalogue spans a large redshift range, up to $z=2.3$, and covers one octant of the sky (close to $5157$ $\mathrm{deg}^2$). 

The \Euclid satellite will observe 15\,000 $\mathrm{deg}^2$ of the sky with patches that extend up to 6000 $\mathrm{deg}^2$. The total area covered by the satellite will be significantly larger than the available Flagship area. By rescaling it is possible to compute the full predicting power from \Euclid. The larger \Euclid survey coverage will allow us to increase statistics, reducing the size of the error bar in particular for the high radius end of the void size function, and to better account for super-sample covariance.
On the other hand, the \Euclid survey is expected to have a less regular pattern than the Flagship box, which might impact the void statistics. Conversely to galaxies, voids are strongly sensitive to survey area specifics because of their extended nature: while contiguous regions are a great advantage for void search, as they provide larger voids, void statistics can be reduced in the case of patchy survey coverage, because voids touching survey edges must be excluded from the analysis.
While the interplay between these different effects may have a role in final constraints, we do not expect this role to significantly impact the precision of constraints resulting from \Euclid. 

We focused our analysis on the expected sub-sample corresponding to spectroscopic data, selecting galaxies from redshift $0.9$ to $1.8$. We obtained a resulting mock catalogue composed of about $6.5 \times 10^6$ galaxies, having the spatial distribution of a shell of sphere octant.

\subsection{Void finding and catalogue preparation}\label{subsec:VIDE} 

We identified cosmic voids in the Flagship light-cone with the public Void IDentification and Examination toolkit\footnote{\url{https://bitbucket.org/cosmicvoids/vide_public}} \citep[\texttt{VIDE},][]{vide}, a parameter-free watershed void finding algorithm based on the code ZOnes Bordering On Voidness \citep[\texttt{ZOBOV},][]{zobov}. \texttt{VIDE} provides a robust density field estimation via the Voronoi tessellation of tracer positions, which allows us to identify local minima and their surrounding density depressions in the tracer density field. With the watershed algorithm \citep{platen2007}, \texttt{VIDE} constructs the void catalogue and provides void properties, such as the void barycentre, the effective radius, the ellipticity, etc. \texttt{VIDE} can  be launched on any catalogue of tracers, both on simulation boxes with periodic boundary conditions and on galaxies from real surveys. It is also capable to handle a survey selection function and a mask. These features make \texttt{VIDE} a very flexible tool to study voids in data and simulations. \texttt{VIDE} has been extensively used for cosmological applications relying on voids in the past decade \citep[see e.g.][]{sutter2012,sutter2014,leclercq2015,hamaus2016,hamaus2017, pollina2019,fang_DES2019, hawken2020, hamaus2020}. 

\begin{table*}
\caption{Void counts measured in the redshift-space mock galaxy catalogue considering the redshift bins and selections used for this analysis. The first column represents the minimum and the maximum redshift values for each bin, while the second and the third columns provide the volume in units of $(h^{-1} \ \mathrm{Gpc})^3$ corresponding to each shell of the sky octant, and the mean separation between galaxies (MGS), respectively. The fourth column reports the factor, $f_{\rm cut}(z)$, used to select voids unaffected by the incompleteness of counts. The last two columns show the number counts of voids identified by the \texttt{VIDE} void finder with $R> f_{\rm cut}(z) \, \mathrm{MGS}$ and of voids obtained after the cleaning procedure with $R_\text{eff}> f_{\rm cut}(z) \, \mathrm{MGS}$, respectively. In the last row we show the total volume of all redshift shells, the mean MGS and $f_{\rm cut}(z)$ values and the total void counts corresponding to the entire range of redshifts. A table with equi-spaced redshift bins is provided in \Cref{AppendixA} to serve as a reference for future forecast analyses needing void numbers.}

\centering
\begin{tabular}{ccccccc}
\toprule
$z$ range & shell volume $[(h^{-1} \ \mathrm{Gpc})^3]$ & MGS $[h^{-1} \ \mathrm{Mpc}]$ & $f_{\rm cut}(z)$ &
all voids & voids after cleaning \\
\midrule
$0.950-1.035$  & $1.157$ & $10.28$ & $2.30$ & $4989$ & $343$   \\
$1.035-1.126$  & $1.329$ & $11.02$ & $2.24$ & $4935$ & $343$   \\
$1.126-1.208$  & $1.269$ & $11.74$ & $2.18$ & $4232$ & $342$   \\
$1.208-1.318$  & $1.796$ & $12.63$ & $2.12$ & $5302$ & $341$  \\
$1.318-1.455$  & $2.363$ & $13.51$ & $2.06$ & $5935$ & $342$   \\
$1.455-1.700$  & $4.490$ & $14.45$ & $2.00$ & $8435$ & $343$   \\
\midrule
$0.950-1.700$  & $12.40$ & $13.69$ & $2.15$ & $33\,828$ & $2054$   \\
\noalign{\vspace{0.01cm}}
\hline 
\bottomrule

\end{tabular}
\label{tab:void_counts}
\end{table*}

We built void catalogues using \texttt{VIDE} from the galaxy sample both with real and redshift-space coordinates given by true and observed redshifts, and note that the redshift-space catalogue is identical to the one used in our companion paper, \citet{EuclidHamaus2021}. In the true redshift catalogue, the galaxy redshift corresponds to the cosmological one only, in the observed redshift catalogue it corresponds to the cosmological plus Doppler shift due to peculiar velocity.

Despite \texttt{VIDE} being a parameter-free algorithm, the theoretical model of the void size function requires voids with the same level of embedded underdensity, so we further processed the void catalogue. We applied therefore to both the obtained void catalogues a cleaning algorithm\footnote{This algorithm is an improved version of the code developed by \cite{ronconi2017} and is inserted in the free software C++/Python libraries \texttt{CosmoBolognaLib V5.5} \citep{CBL}, available at \url{https://gitlab.com/federicomarulli/CosmoBolognaLib}. In this version of the code, the cleaning procedure can be applied to catalogues with comoving coordinates and the void rescaling is performed by taking into account the variation of the tracer density with redshift.} \citep{ronconi2017}. The goal of this procedure is to conform observed voids with their theoretical counterpart. The main steps of this cleaning pipeline are: (i) the rejection of spurious voids, i.e. with central density too high or with radius below the spatial resolution of the tracer catalogue, (ii) the rescaling of voids to a specific radius $R_{\rm eff}$ to match a specific spherical density contrast within the sphere, $\delta_\mathrm{v,tr}^\mathrm{NL}$, in the tracer distribution, (iii) the removal of overlapped voids, i.e. voids whose distance between centres is smaller than the sum of their radii.
We underline that, during the rescaling procedure, any negative value of density contrast $\delta_\mathrm{v,tr}^\mathrm{NL}$ can in principle be chosen to resize underdensities, as long as the theoretical model is consistently calculated using the same threshold (see \Cref{subsec:theory_Vdn}). When dealing with observed voids, the threshold can be fixed to a suitable value chosen based on survey features.
We considered the following reasoning to select this value: on the one hand the more negative the threshold, the more the identified underdensities are free of contamination by Poisson noise \citep[see also][for a discussion on spurious voids and possible treatments]{zobov,cousinou2019} and the stronger the impact of the cosmology on the void size function; on the other hand, an excessively negative threshold entails both a low statistic and a higher uncertainty in the rescaled void radius, caused by the sparsity of galaxies tracing such extreme underdense regions.
In particular, for this analysis we followed the choice of \cite{contarini2019,Contarini2020}, selecting a threshold $\delta_\mathrm{v,tr}^\mathrm{NL}=-0.7$, which ensures a good compromise on the aforementioned effects.
We verified the robustness of our method by also performing the entire analysis using $\delta_\mathrm{v,tr}^\mathrm{NL}=-0.6$, finding consistent results.

\texttt{VIDE} takes into account the presence of a survey mask, and prevents voids from including volumes outside the survey extent.
We applied the mask following the simulated $\sim5000 \ \mathrm{deg}^2$ octant. 
While the actual \Euclid data will be more complex (due to e.g. more elaborate survey mask and survey-related systematic effects), this methodology at least partially accounts for mask effects in our pipeline, preparing the analysis of future \Euclid data.
Aiming at a very conservative void selection at the edges of the survey's footprint, we applied an additional cut to ensure the mask is not affecting the cleaning procedure: we removed all voids whose centre is closer than $30 \ h^{-1} \ \mathrm{Mpc}$ to the edge and corrected
the model accordingly for the selected volume.
We then pruned voids at low and high redshifts to further avoid selection effects given by redshift boundaries of the light-cone, and we divided the sample in six redshift bins. This number is found as the optimal compromise between maximising the number of redshift shells and keeping void numbers in bins high enough to avoid falling in the shot-noise dominated regime. In order to have shells with roughly the same number of cleaned voids identified in redshift space and to avoid border effects at the light-cone redshift boundaries, we selected the following redshift bin edges: $z_i = [0.950, 1.035, 1.126, 1.208, 1.318, 1.455, 1.700]$.
Each shell contains at least $340$ voids, within the range of effective radii considered in the analysis of the measured void size function described below. 

Tracer sparsity leads to a drop of counts for small voids in the measured void size function \citep{Sutter2013,verza2019}. The incompleteness depends on the mean galaxy separation and therefore on the redshift of the sample \citep{jennings2013, ronconi2019, contarini2019, verza2019}. Modelling the drop of counts for small voids is not trivial. To avoid falling in this regime, we conservatively excluded from the analysis voids with radii falling in the range of scales affected by incompleteness. We removed voids with radii smaller than $\mathrm{MGS} \, f_{\rm cut}(z)$, where MGS is the mean galaxy separation and $f_{\rm cut}(z)$ is a factor dependent on the redshift of the sample. 
We computed the value of mean galaxy separation as $\mathrm{MGS} = (V_\mathrm{shell}/N_\mathrm{gal})^{1/3}$, where $V_\mathrm{shell}$ is the volume of the redshift shell analysed and $N_\mathrm{gal}$ is the number of galaxies present in it.
The factor $f_{\rm cut}(z)$ is chosen empirically based on the drop of void counts and on the steep departure from the theoretical model. We found that values spanning from $2.3$ (lowest redshift bin) to $2$ (highest redshift bin) for $f_{\rm cut}(z)$ ensure the exclusion of spatially unresolved voids in redshift space. 

Since we expect the resulting void size function in redshift space to be shifted towards greater effective radii due to the effects of RSD \citep{pisani2015b, zhao2016, nadathur2016b, Correa2020}, we extended the minimum radius for the real-space case, adding an extra bin at small radii while keeping the same binning of the redshift-space case for higher bins. We verified that these choices allow us to be outside of the incompleteness regime, for both the void size function in real and redshift space. 

In \Cref{tab:void_counts} we show the number counts of voids selected from the redshift-space mock galaxy catalogue. For each of the redshift bins with edges $z_i$ we report the volume occupied by the shell and the MGS of the tracers, together with the factor $f_{\rm cut}(z)$ used to compute the minimum void radius considered in this analysis. For completeness, we show the void number counts both before and after the cleaning procedure aimed to line up observed voids to theoretical voids, according to the void size function model. The sharp decrease of the void number is an expected outcome of the cleaning procedure, that selects the largest in volume and deepest underdensities identified by \texttt{VIDE} and rescales their sizes towards smaller values, causing a more severe rejection of voids during the removal of the spatial scales affected by the incompleteness of counts. Although this conservative approach leads to a loss of the void size function constraining power, it ensures the selection of an high-purity void sample and a robust treatment of void number counts.
In future works, different approaches will be explored to improve the void selection also at small radii: among these, the application of machine learning techniques \citep{cousinou2019} is promising to carefully remove only spurious voids and consequently enhance the performance of the void size function as a cosmological tool.

\section{Theory and methods}\label{sec:theory_&_methods}

In this section we introduce the theoretical background necessary for this work. We first discuss the model of the void size function, then we present the prescriptions applied to extend this model to voids identified in the distribution of biased tracers. We describe the Bayesian statistical analysis used to provide forecasts on the DE equation of state and on the sum of neutrino masses. Finally, we present the cosmological scenarios considered in our analysis.

\subsection{Theoretical void size function}\label{subsec:theory_Vdn}
To estimate the constraining power of the void size function, i.e. the distribution function of void radii, we first need a theoretical model. The void size function model most widely used in the literature relies on the excursion-set formalism, developed within the framework of the halo mass function \citep{peacock_heavens1990,cole1991,bond1991,mo_white1996}. This model was first  proposed by \citet{SVdW2004} and extended by \citet{jennings2013}. The distribution of fluctuations that become voids, i.e. the multiplicity function, is obtained as the conditional first crossing distribution of the matter density contrast filtered at decreasing Lagrangian radius in a double barrier problem: a fluctuation becomes a void at a radius $R_\mathrm{v}$ if the filtered density contrast first crosses the void formation threshold $\delta_\mathrm{v}^\mathrm{L}$ at $R_\mathrm{v}$, without having crossed the threshold for collapse $\delta_\mathrm{c}^\mathrm{L}$ at any larger scale\footnote{In this paper the density contrasts derived in linear and nonlinear theory are indicated with the superscripts $\mathrm{L}$ and $\mathrm{NL}$, respectively. In absence of any superscript, we take for granted the reference to the nonlinear counterpart.} \citep{SVdW2004}. The multiplicity function of \cite{SVdW2004} is derived for spherical fluctuations in Lagrangian space, i.e. the initial density field linearly evolved to the epoch of interest, while the observed voids live in the fully nonlinear evolved density field in comoving coordinates, i.e. the Eulerian space. The spherical approximation allows us to easily go back and forth from Lagrangian to Eulerian space in all the computations.

The void size function probes the inner region of cosmic voids and in contrast to the collapsing case, i.e. halo formation \citep{monaco1995,sheth_tormen2002}, the spherical approximation is accurate enough for this purpose, at least for voids of scales detectable by \Euclid \citep{icke1984,verza2019}.

The linear threshold for collapse is fixed at $\delta_\mathrm{c}^\mathrm{L}=1.686$, according to the collapse of a spherical fluctuation. This value corresponds in an Einstein--de Sitter model to the full collapse in linear theory, when the halo virialises. 
The void case is different: if the initial underdensity identifying a void is deep enough, its evolution is not marked by any specific event, and it continues its outward-directed expansion forever. 
It is common to consider the shell-crossing\footnote{In a purely theoretical framework, cosmic voids can be represented by negative perturbations modelled as a set of concentric shells. During their evolution, the inner shells will expand faster than the outer ones and will eventually overtake the more external ones, giving rise to the so-called `shell-crossing' phenomenon. During this event the trajectories of two fluid elements in Lagrangian coordinates cross each other, breaking the one-to-one map between Lagrangian and Eulerian space. See \citet{massara2018} for a detailed study on the mapping between Lagrangian and Eulerian voids.} condition as the event that identifies the void formation \citep{blumenthal1992,SVdW2004,jennings2013}, but this condition strictly depends on the initial density profile of the  underdensity. For an initial density profile represented as a step function, shell-crossing happens at a nonlinear matter density contrast of  $\delta^\mathrm{NL}_\mathrm{sc} \simeq -0.8$, corresponding to a linear threshold of $\delta_\mathrm{sc}^\mathrm{L} \simeq -2.7$. Considering more physical density profiles \citep[e.g.][]{hamaus2014,massara2018}, shell-crossing in voids does not necessarily happen and, if it does, it may occur at even lower threshold values.  
Given the considered thresholds, our voids remain far from the shell-crossing regime, therefore, in principle, it is always possible to map the measured Eulerian density profile to the corresponding Lagrangian one.
As we do not reach shell-crossing, we have the freedom to choose any threshold value to define void formation \citep{ronconi2019, contarini2019, verza2019, Contarini2020}.

The multiplicity function, as given by \cite{SVdW2004}, is:
\begin{equation}
f_{\ln \sigma}(\sigma) = 2 \sum_{j=1}^{\infty} \, \exp{\bigg(-\frac{(j \pi x)^2}{2}\bigg)} \, j \pi x^2 \, \sin{\left( j \pi \mathcal{D} \right)}\, ,
\end{equation}
with
\begin{equation}
\mathcal{D} = \frac{|\delta_\mathrm{v}^\mathrm{L}|}{\delta_\mathrm{c}^\mathrm{L} + |\delta_\mathrm{v}^\mathrm{L}|}\, , \qquad x = \frac{\mathcal{D}}{|\delta_\mathrm{v}^\mathrm{L}|} \sigma \, ,
\end{equation}
where $\sigma$ is the square root of the variance of linear matter perturbations on the Lagrangian scale $r_{\rm L}$. All these quantities are computed in the linear regime, on which the excursion-set formalism relies. The void size function in Lagrangian space is then readily derived as \citep{SVdW2004,jennings2013}:
\begin{equation}\label{liearVdn}
\frac{\diff n_\mathrm{L}}{\diff \ln r_\mathrm{L}} = \frac{f_{\ln \sigma}(\sigma)}{V(r_\mathrm{L})} \, \frac{\diff \ln \sigma^{-1}}{\diff \ln r_\mathrm{L}}\,,
\end{equation}
where $V(r_\mathrm{L})=4 \pi r_\mathrm{L}^3 /3$ is the volume of the spherical fluctuation of radius $r_\mathrm{L}$. 
Conversely to the case of the halo mass function, the void size function in Eulerian space is different from the one in Lagrangian space. Firstly, the expansion of voids from linear to nonlinear theory has to be taken into account. The evolution of perturbations in the nonlinear regime provides the conversion from the linear to the nonlinear shell radius:
\begin{equation}\label{r_r_L}
\frac{r}{r_\mathrm{L}} = \left( \frac{\overline{\rho}}{\rho_\mathrm{v}} \right)^{1/3}\, ,
\end{equation}
where $\overline{\rho}$ is the mean density of the Universe and $\rho_\mathrm{v}$ is the average density within the void. Secondly, 
to prevent the fraction of the volume occupied by voids from exceeding unity in the transition from linearity to nonlinearity, we fix the void volume fraction of the Universe to be equal in the linear and in the nonlinear regimes \citep{jennings2013}:
\begin{equation}
V(r) \ \diff n = V(r_\mathrm{L}) \, \diff n_\mathrm{L} \rvert_{r_\mathrm{L} = r_\mathrm{L}(r)}\, .
\end{equation}
With this requirement the model ensures void volume conservation -- hereafter Vdn model, following \cite{jennings2013} -- and from \Cref{liearVdn} we can derive the final definition of the theoretical void size function adopted in this paper:
\begin{equation}
\frac{\diff n}{\diff \ln r} = \frac{f_{\ln \sigma}(\sigma)}{V(r)} \, \frac{\diff \ln \sigma^{-1}}{\diff \ln r_\mathrm{L}} \biggr \rvert_{r_\mathrm{L} = r_\mathrm{L}(r)}\, .
\end{equation}

\subsection{Methodology}\label{subsec:methodology}


To compare the measured and the theoretical void size functions, we need to link objects found by the void finder in the tracer distribution with the ideal spherical and isolated voids described by the void size function theoretical model \citep{jennings2013}. Any watershed void finder defines a region spanning from its density minimum to its overdense ridge
\citep{roerdink2000,zobov,platen2007,vide}. On the contrary, the theoretical voids are matter density fluctuations for which the mean density contrast in a sphere reaches a specific threshold value at a radius $R_\mathrm{eff}$. 
Previous papers attempted to mitigate this difference by modifying the threshold of the model \citep{pisani2015,sahlen2016,sahlen2019}, in particular considering marginalisation over the threshold, for cosmological uses of the model.

\begin{figure*}
\centering
\includegraphics[width=0.75\textwidth]{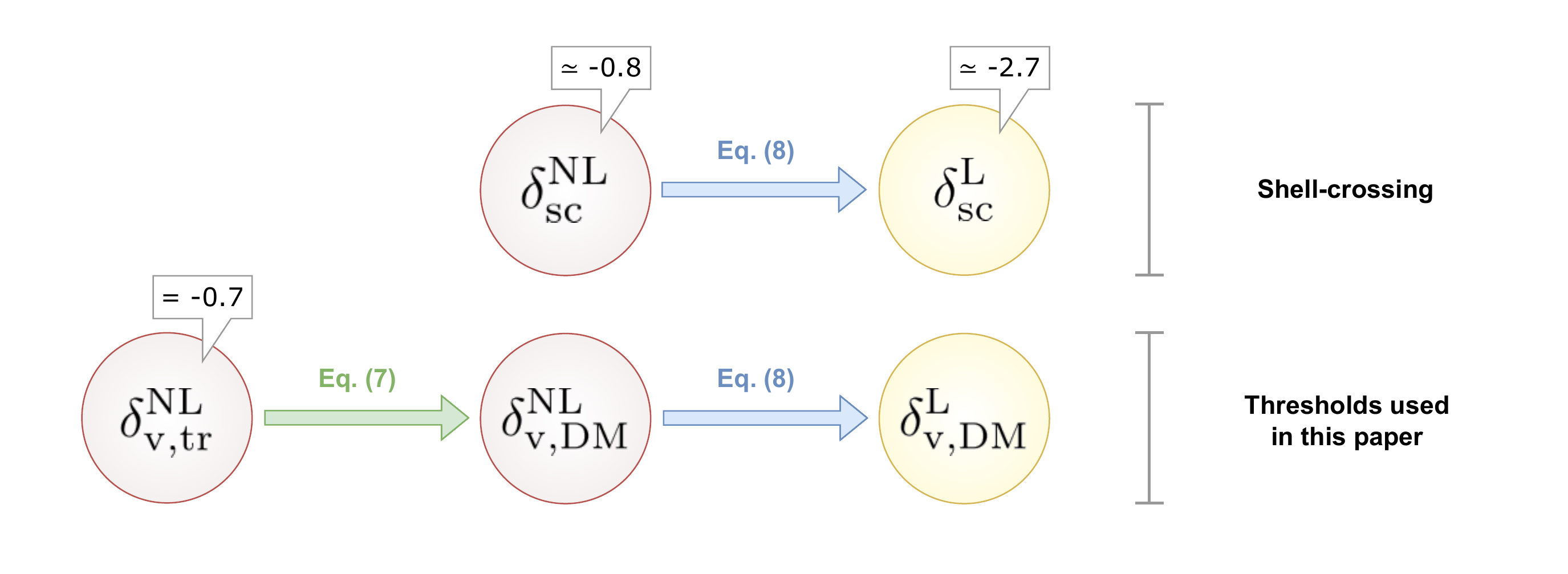} 
\caption{Summary of the underdensity thresholds and corresponding relationships introduced in this work. In the first row we report the values relative to the shell-crossing phenomenon (subscript $\mathrm{sc}$) together with their corresponding numerical values, shown in the white tags. In the second row we list the density contrasts values used in our analysis. The quantities circled in red refer to density contrasts defined in nonlinear theory (superscript $\mathrm{NL}$), while those circled in yellow are defined in linear theory (superscript $\mathrm{L}$). The latter are used to compute the Vdn model and are obtained from the former by means of \Cref{eq:bernardeau}. All the quantities listed in the central and in the right columns are defined in the density field traced by unbiased objects, i.e. DM particles (subscript $\mathrm{DM}$). On the far left we report instead the underdensity threshold used to define void in this paper. It has a value of $-0.7$ and is computed in the density field traced by galaxies (subscript $\mathrm{tr}$). To find its corresponding value in the underlying DM field we make use of \Cref{thr_conversion}.} 
\label{fig:thresholds}
\end{figure*}

It is useful to recall that the Vdn model describes voids evolving in the total matter density field, but that in our case (and when dealing with data) we can only identify voids in the galaxy density field. Therefore, to align these objects to those modelled by the theory, we need to relate the characteristic density threshold used in the theoretical model, $\delta_\mathrm{v}^\mathrm{L}$, to the corresponding one in the galaxy density field.
To accomplish this purpose, we relied on the following two steps for data preparation: first, we measured the mean density profiles of cosmic voids to find the radius of the sphere at which the mean density contrast reaches the desired value $\delta_\mathrm{v,tr}$ in the galaxy distribution \citep{jennings2013,ronconi2017,ronconi2019,contarini2019,verza2019}, i.e. the resized radius, $R_\mathrm{eff}$.
Second, it is necessary to find the corresponding density contrast in the underlying matter density distribution, within the resized radius.
Recently the properties of voids in the galaxy distribution, as well as of galaxies and tracer bias within cosmic voids, have been explored extensively \citep{furlanetto2006,Sutter2013,Neyrinck_2014,pollina2017,pollina2019,contarini2019,Contarini2020}. 
To recover the matter density contrast corresponding to the threshold value in the galaxy density field, we need to model the galaxy distribution inside cosmic voids taking into account tracer bias.
To describe tracer bias, i.e the bias of the cosmological objects chosen to trace voids, various possibilities have been considered, including a full theoretical description (see discussion in \citealp[]{verza2019} and \citealp[]{desjacques_bias2016} for an extensive review), or a robust modelling of bias inside voids based on simulations \citep{pollina2017, pollina2019, contarini2019, Contarini2020}. In this analysis we chose to rely on the latter, following \citet{contarini2019, Contarini2020}. 
These works showed that it is possible to extend the Vdn model by considering a linear relationship between tracer and matter density contrast in cosmic voids,
$\delta_\mathrm{v,tr}^\mathrm{NL}$ and $\delta_{\mathrm{v},\mathrm{DM}}^\mathrm{NL}$, with a dependence only on the large-scale effective bias $b_\mathrm{eff}$:
\begin{equation}\label{thr_conversion}
\delta_\mathrm{v,DM}^\mathrm{NL} = \frac{\delta_\mathrm{v,tr}^\mathrm{NL}}{\mathcal{F}(b_\mathrm{eff})}\, ,
\end{equation}
where $\delta_\mathrm{v,DM}^\mathrm{NL}$ is the value of the threshold in the dark matter field to be used in the Vdn model, after its conversion in linear theory \citep{jennings2013}. For $\Lambda$CDM and the DE equations of state considered in this work, the conversion from nonlinear to linear density contrast in the matter field is cosmology and redshift independent with very high accuracy \citep{jennings2013,Pace2017}, allowing us to exploit the fast and precise \citet{bernardeau1994} fitting formula:
\begin{equation} \label{eq:bernardeau}
  \delta_\mathrm{v}^\mathrm{L} = \mathcal{C} \, \bigl[1 - (1 + \delta_\mathrm{v}^\mathrm{NL})^{-1/\mathcal{C}}\bigr] \ , \ \text{with } \mathcal{C}=1.594 \, .
\end{equation}
Comparing the void density profiles computed both in the Friend-of-Friends (FoF) halos and dark matter particle field, \citet{contarini2019} found that the function $\mathcal{F}$, reported in \Cref{thr_conversion}, is well modelled as a linear relation of the large-scale effective bias $b_\mathrm{eff}$:
\begin{equation}\label{b_punct_formula}
\mathcal{F}(b_\mathrm{eff}) = B_\mathrm{slope} \, b_\mathrm{eff} + B_\mathrm{offset} \, ,
\end{equation}
where $B_\mathrm{slope}$ and $B_\mathrm{offset}$ are the values of the first and second coefficients of the linear function, respectively. This relation will be calibrated in this work using the samples of galaxies and voids extracted from the Flagship light-cone and the resulting values of $B_\mathrm{slope}$ and $B_\mathrm{offset}$, together with the associated uncertainties, will be presented in \Cref{subsec:Vdn_analysis}. The parametrisation introduced in equations \Cref{thr_conversion,b_punct_formula} was tested also in \citet{Contarini2020}, using different selection criteria for the halo identification and verifying its negligible dependence on the cosmological model. In particular, this relation is tested varying the neutrino mass and the parameters of the $f(R)$ class of modified gravity models, in the form introduced by \cite{HuSawicki2007}. The quantity represented by the function $\mathcal{F}(b_\mathrm{eff})$ parametrises the value of the tracer effective bias measured inside cosmic voids and it has been denoted $b_\mathrm{punct}$\footnote{We note that $b_\mathrm{punct}$ and $\mathcal{F}(b_\mathrm{eff})$ refer to the same quantity, i.e. the value of the tracer bias computed inside cosmic voids. This quantity represents the relation between the void density profiles computed using biased (e.g. galaxies) and unbiased (e.g. dark matter particles) mass tracers \citep[see][]{pollina2017,pollina2019,contarini2019,Contarini2020}. Nevertheless, we keep two different notations since their computation is different: $b_\mathrm{punct}$ is measured for each bin of redshift, while  $\mathcal{F}(b_\mathrm{eff})$ is given as a function of $b_\mathrm{eff}$ and varies linearly with it by construction (see \Cref{subsec:Vdn_analysis}).}
in \citet{contarini2019,Contarini2020}.
To facilitate the reader's comprehension of the adopted methodology, we summarise in \Cref{fig:thresholds} all the negative density contrasts mentioned in this paper and their relative relations.

To convert the underdensity threshold of the Vdn model according to the function $\mathcal{F}(b_\mathrm{eff})$, we first need to compute  accurately the large-scale effective linear bias of our galaxy sample. For this estimate we followed the same prescriptions described in \cite{marulli2013} and \cite{Marulli2018}. In particular, we exploited the galaxy two-point correlation function (2PCF), performing a Bayesian statistical analysis to infer the effective bias, $b_{\text{eff}}$.
We computed the angle-averaged galaxy 2PCF $\hat{\xi}(r)$ in real space creating a random catalogue $10$ times larger than the original one and using the Landy \& Szalay estimator \citep{Landy_Szalay1993}.

We then estimated the covariance matrix, which measures the variance and correlation between the different bins of the 2PCF. For this purpose we applied the Bootstrap method, dividing the original catalogues in 125 sub-catalogues and constructing 100 realisations by resampling from the sub-catalogues, with replacement. In the end we performed a full Markov chain Monte Carlo (MCMC) analysis of the 2PCF, using a Gaussian likelihood function. The 2PCF model, $\xi_\mathrm{mod}(r)$, is computed as follows:
\begin{equation}
\xi_\mathrm{mod}(r)= b^2_{\text{eff}} \, \xi_{\rm m}(r) \, ,
\end{equation}
where $\xi_{\rm m}(r)$ is the matter 2PCF, which is estimated by Fourier transforming the matter power spectrum, $P_\mathrm{m}(k)$, computed with the {\small Code for Anisotropies in the Microwave Background} \citep[\texttt{CAMB}\footnote{\url{http://camb.info}},][]{LewisCAMB}.
Then we accurately estimated the effective bias parameter $b_{\text{eff}}$ by sampling its posterior distribution with the MCMC modelling in the range of scales of $[20-40] \ h^{-1} \ \mathrm{Mpc}$. 

We underline that the relative error associated to $b_{\text{eff}}$ is expected to be relatively small because of the strategy used to compute this quantity relying on the galaxy catalogue in real space and assuming the true cosmological parameters of the simulation. A more complete and realistic treatment will be performed in the future, including in the analysis the modelling of the multipoles of the 2PCF, which will allow us to take into account the effects of redshift-space and geometrical distortions \citep[see e.g.][]{Scocimarro2004,Taruya2010,Beutler2017,Pezzotta2017}.

We finally recall that another approach to compute effective bias, analogous to that applied in this work, is to measure the 2PCF in Fourier space and to model it via the theoretical matter power spectrum $P(k)$ \citep[see e.g.][]{Beutler2017}. Additionally, an alternative methodology to extract Flagship galaxy bias is to follow e.g. \citet{Tutusaus2020}, who parametrised the Flagship galaxy bias as a function of $z$, albeit for the photometric redshift selection.

\subsection{Bayesian statistical analysis}\label{subsec:forecasts_theory}

In this work we used a reliable forecast method for the sensitivity of the void size function in the \Euclid survey to constrain the cosmological model, based on a parameter extraction from Bayesian analysis with MCMC \citep{perotto2006,wang2009,lahav2010,martinelli2011,debernardis2011,wolz2012,hamann2012,khedekar2013,audren2013}.

In order to forecast the sensitivity of void counts with an MCMC analysis in \Euclid, we have to consider that the Flagship simulation covers about one third of the \Euclid survey. We obtained the \Euclid predicted void number counts relying on the theoretical void size function model validated on the Flagship simulation (see \Cref{subsec:Vdn_analysis}), that is assuming a fiducial \lcdm cosmology with the cosmological parameters of the Flagship and the calibration in redshift space of the Vdn model described in \Cref{subsec:Vdn_analysis}. We assumed the same binning of void radii employed in our Flagship analysis but consider a survey area matching the one expected for \Euclid (roughly three times the Flagship area), rescaling the Poissonian errors of the void number counts consistently by a factor $\sqrt{3}$.

This allows us to use MCMC analysis to explore the likelihood distribution in the parameter space without any assumption on the Gaussianity of parameters and local approximations around the fiducial value, as in Fisher forecasts. Moreover, according to the Cramér--Rao inequality, the Fisher matrix gives a lower bound on the error on a parameter \citep{Kendall_statistics}, while the MCMC is proven to be more realistic, in particular in the presence of degeneracies \citep{perotto2006,wolz2012,audren2013,Sellentin2014}. Finally, this kind of approach allows us to compute unbiased constraints, with confidence contours centred on the Flagship simulation cosmological parameters and on the calibrated nuisance parameters $B_\mathrm{slope}$ and $B_\mathrm{offset}$.

According to Bayes's theorem, given a set of data $\mathcal{D}$, the distribution of a set of parameters $\Theta$ in the cosmological model considered is given by the posterior probability:
\begin{equation}
\mathcal{P}(\Theta | \mathcal{D})  \propto \mathcal{L} (\mathcal{D} | \Theta) \, p(\Theta)\,,
\end{equation}
where $\mathcal{L} (\mathcal{D} | \Theta)$ is the likelihood and $p(\Theta)$ the prior distribution. Since in this work we consider the number counts of cosmic voids, the likelihood can be assumed to follow Poisson statistics \citep{sahlen2016}:
\begin{equation}
\mathcal{L} (\mathcal{D} | \Theta) = \prod_{i,j} \frac{N(r_i,z_j | \Theta)^{N(r_i,z_j | \mathcal{D})} \, \exp{\left[ -N(r_i,z_j | \Theta)\right]}}{N(r_i,z_j | \mathcal{D})!} \,,
\end{equation}
where the product is over the radius and redshift bins, labelled as $i$ and $j$ respectively.  The $N(r_i,z_j | \mathcal{D})$ quantity corresponds to the number of voids in the $i^\mathrm{th}$ radius bin and $j^\mathrm{th}$ redshift bin, while $N(r_i,z_j | \Theta)$ corresponds to the expected value in the cosmological model considered, given a set of parameters $\Theta$. 
In our work, the former is obtained from the Flagship analysis (with the void size function model validated on the Flagship simulation, but considering that the \Euclid area will be three times larger), while the latter is given by the predictions of the void size function model varying the considered cosmological parameters $\Theta$.

In performing the MCMC analysis, the mapping between redshift and comoving distance changes with the cosmological parameters assumed at each step of the chain. This introduces geometrical distortions for all the considered sets of cosmological parameters (different from the true one). We used a fiducial cosmology to build up the void catalogue, and, in computing the likelihood, we theoretically accounted for the distortion effects on the quantities we measured. In particular, geometrical distortions can be modelled with two effects: they vary the inferred survey comoving volume and introduce the Alcock--Paczy{\'n}ski \citep{AP1979} distortion. The effect on the survey volume impacts the number of voids expected in the survey. The theoretical void size function model predicts the number density of voids in each radius and redshift bin. Therefore, to obtain the total number of voids, the number density has to be multiplied by the volume, which is impacted by the cosmology.
On the other hand, the Alcock--Paczy{\'n}ski \citep{AP1979} distortion affects the size of voids and introduces an anisotropy between the orthogonal and the parallel direction with respect to the line-of-sight. These quantities change according to \citep{Sanchez2017b}:
\begin{equation}
r_\parallel' = \frac{H(z)}{H'(z)} \, r_\parallel = q^{-1}_\parallel \, r_\parallel \,, \qquad r_\perp' = \frac{D_{\rm A}'(z)}{D_{\rm A}(z)} \, r_\perp = q^{-1}_\perp \, r_\perp \,;
\end{equation}
where $r_\parallel$ and $r_\perp$ are the comoving distances between two objects at redshift $z$ projected along the parallel and perpendicular direction with respect to the line-of-sight, $H(z)$ is the Hubble parameter and $D_{\rm A}(z)$ the comoving angular-diameter distance. The primed quantities refer to the calculation at the fiducial cosmology, the non-primed to the true cosmology, assumed in a MCMC step. It follows that the volume of a sphere with radius $R$ appears modified according to $R=q_\parallel^{1/3}q_\perp^{2/3} R'$ \citep{Ballinger1996, Eisenstein2005_BAO, xu2013, Sanchez2017b, hamaus2020, Correa2020}, so the void size function expected in the survey is shifted accordingly.
We checked the validity of this relationship varying the cosmology used to get the comoving distances from redshifts and consequently correcting the radius $R_\mathrm{eff}$ at which voids reach the underdensity threshold $\delta_\mathrm{v,tr}^\mathrm{NL}$.

We assumed the void centres to remain at the same locations at different cosmologies. While void shapes can suffer from symmetric geometrical distortions, this marginally affects the identification of void centres, and the effect is even smaller since the void size function is an averaged quantity. Furthermore, the variation caused by the change of the cosmological parameters on void radii is taken into account by the modelling of the Alcock--Paczy{\'n}ski effect, therefore the cleaning procedure (see \Cref{subsec:VIDE}) is applied only once to the void sample, considering a fiducial \lcdm cosmology.
We note that the combination of the two effects -- volume effect acting on the expected number density, and the Alcock--Paczy{\'n}ski effect acting on the void sizes -- enhances the constraining power of the void size function.

\begin{figure*}
\centering
\includegraphics[width=0.42\textwidth]{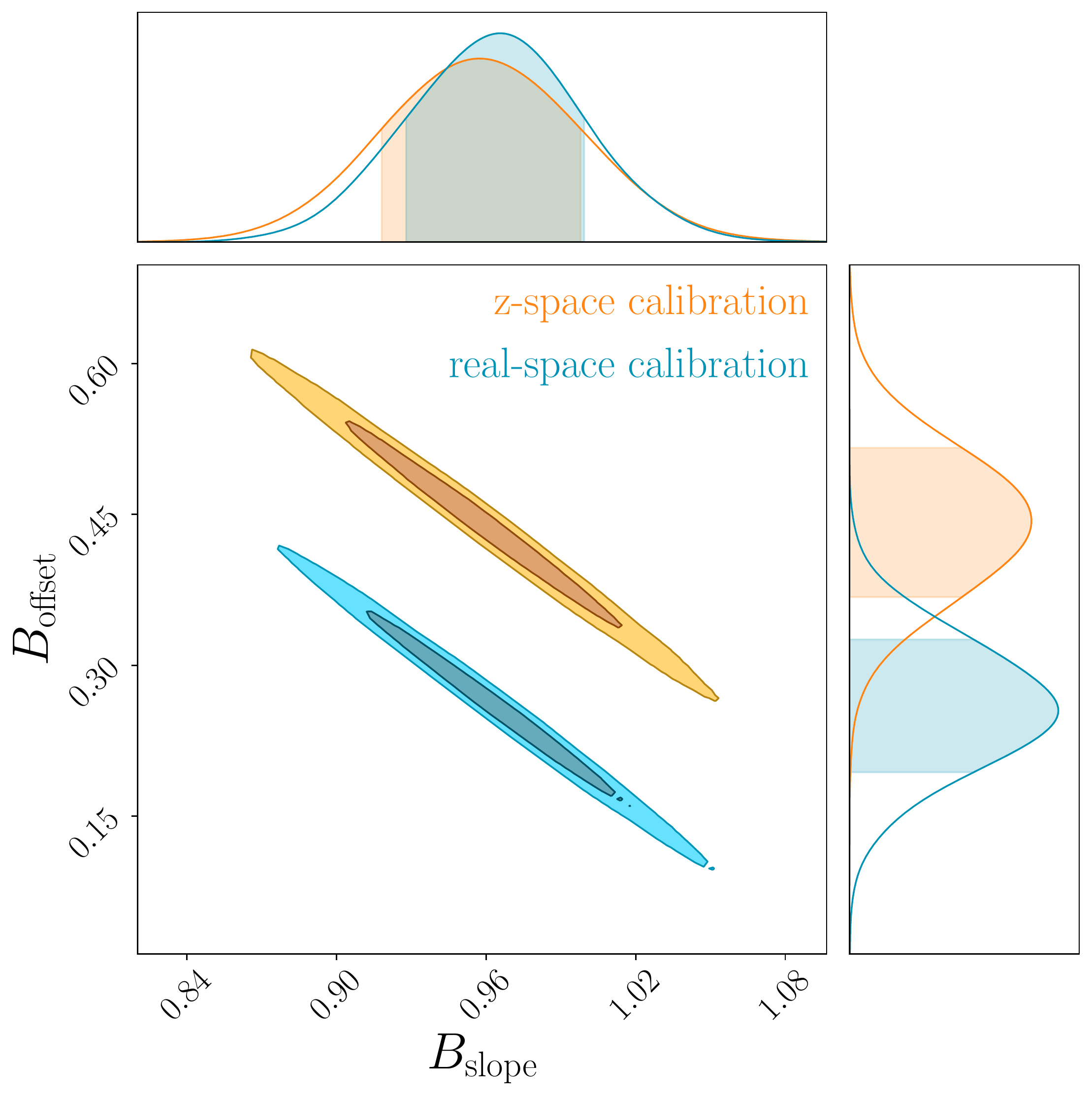} 
\hfill
\includegraphics[width=0.52\textwidth]{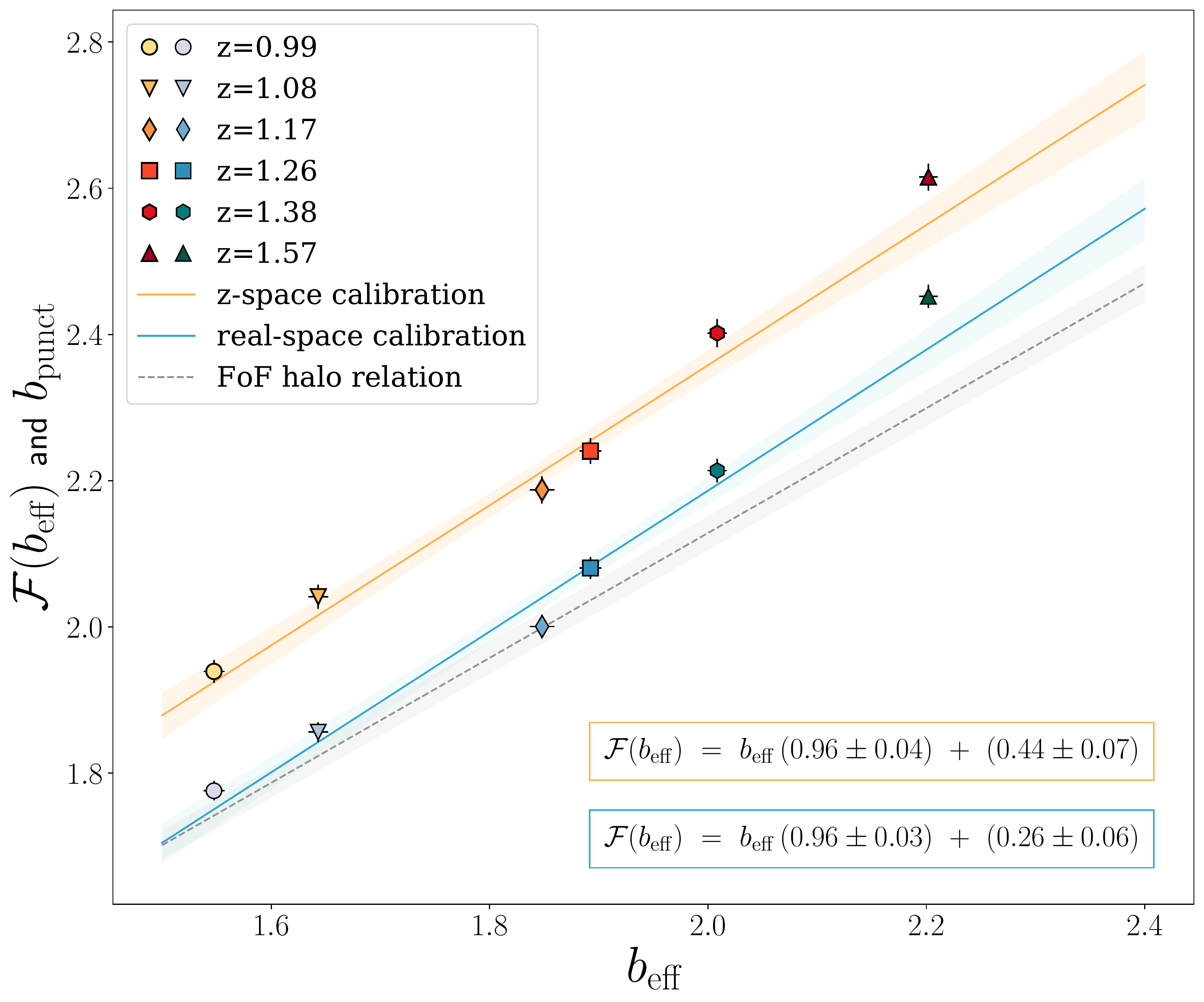}
\caption{Calibration of the relation $\mathcal{F}(b_\mathrm{eff})$ from \Cref{b_punct_formula}, required for the conversion of the threshold $\delta_\mathrm{v,tr}$ in \Cref{thr_conversion}. \textit{Left}: the $68\%$ and $95\%$ confidence levels in the $B_\mathrm{slope}$--$B_\mathrm{offset}$ plane for the void catalogues built both in real (blue) and in redshift space (orange). \textit{Right}: the solid lines represent the resulting linear relations $\mathcal{F}(b_\mathrm{eff})$ obtained with the calibrated coefficients $B_\mathrm{slope}$ and $B_\mathrm{offset}$ for real (blue) and redshift space (orange), while the shaded regions indicate an uncertainty of 2$\sigma$ on the relationships. The markers represent the calibration obtained for each bin of redshift, leaving $b_\mathrm{punct}$ as the only free parameter of the void size function model when fitting the measured void number counts. This alternative calibration provides a value of $b_\mathrm{punct}$ for each redshift of the sample and is associated with the value of the effective bias $b_\mathrm{eff}$ of the Flagship galaxies at that specific redshift. As a comparison we also show the linear function calibrated using FoF dark matter haloes in real space by \cite{contarini2019}, displayed with a dashed grey line.} 
\label{fig:bias_relation}
\end{figure*}

\begin{figure*}
\centering
\includegraphics[width=0.7\textwidth]{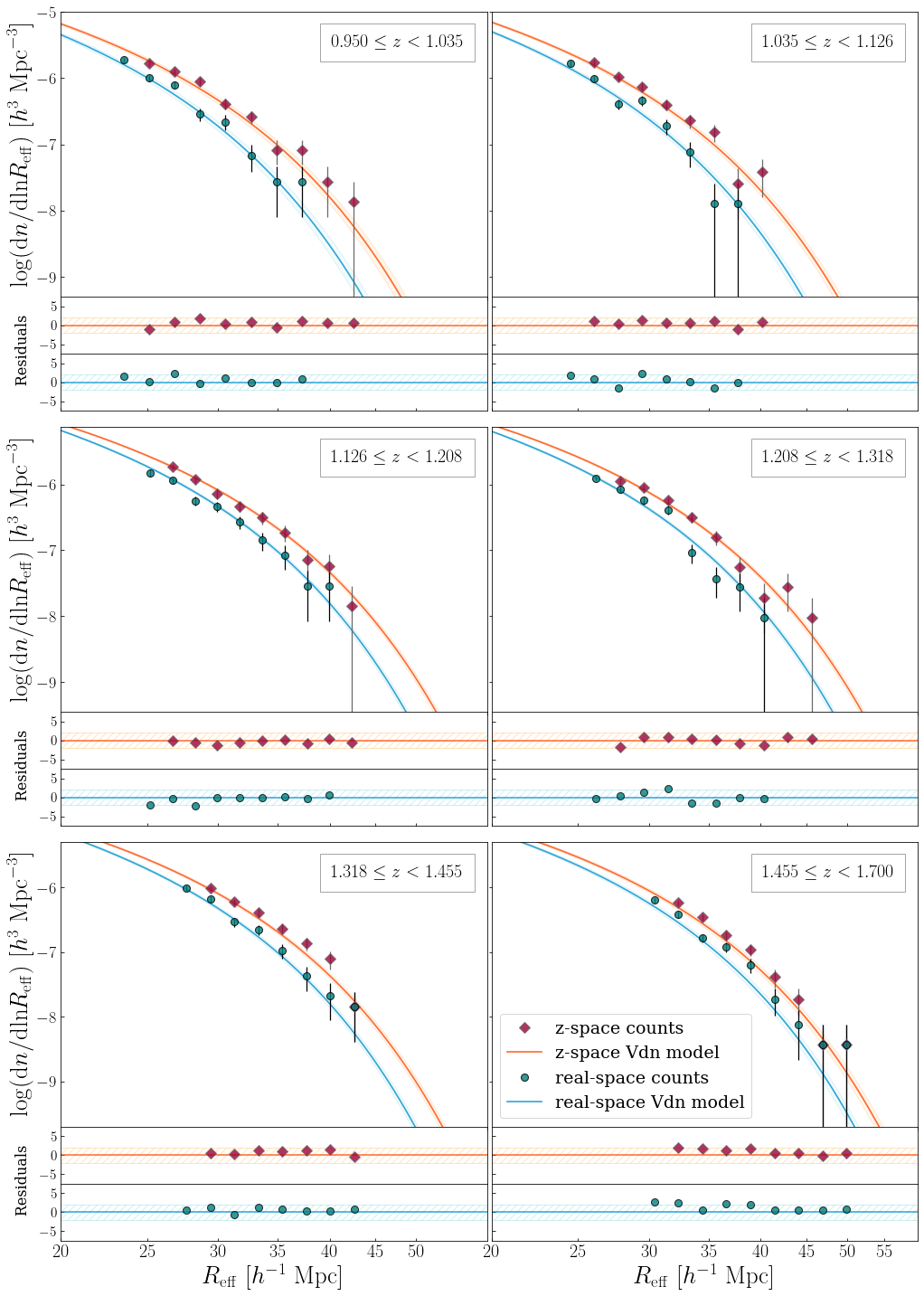}
\caption{Comparison between the measured void number counts as a function of $R_\mathrm{eff}$ (the void radii rescaled by the cleaning algorithm), and the theoretical predictions given by the extended Vdn model, in six different redshift bins. The dark green circles and the dark red diamonds represent the measured void size functions in real and redshift space, respectively, while the corresponding model predictions are depicted in light blue and orange. The shaded regions indicate the uncertainty of $2\sigma$ assigned to the model through the calibration of the extended Vdn parameters. Bottom panels report the residuals computed as the difference of data points from the relative theoretical model, divided by the Poissonian error associated with each data point. The hatched regions represent a band with amplitude $2$ useful to check if the data points, considered with a $2\sigma$ error, are compatible with the main theoretical curve.}
\label{fig:void_size_functions}
\end{figure*}

\subsection{Cosmological models}\label{subsec:cosmological_models}
The aim of this work is to investigate the constraining power of the void number count statistic on cosmological parameters, focusing in particular the DE equation-of-state parameters.
We considered two cosmological models, extending the standard $\Lambda$CDM with different DE equation of states.
The first model, $w$CDM, implements a constant DE equation of state $w$; the second one, $w_0 w_a$CDM, parametrises dynamical DE models with the popular Chevallier--Polarski--Linder (CPL) equation of state \citep{Chevallier&Polarski2001, Linder2003}:
\begin{equation}
    w_\mathrm{CPL}(z) = w_0 + w_a \frac{z}{z+1}\, \text{.}
\end{equation}
Both cosmological models consider a flat universe and do not include spatial fluctuations of the DE, which are negligible given the scales considered in this work \citep[see e.g.][]{Khoraminezhad2020}. We performed the MCMC analysis of each cosmological model focusing on different sets of free cosmological parameters: together with the DE equation of state parameters (i.e. $w$ or $ w_0$ and $w_a$, depending on the cosmological model) the density parameter $\Omega_{\rm m}$ or the sum of neutrino masses $M_\nu$ are allowed to vary. Moreover, we analysed both the cases with two different approaches:
firstly, fixing the parameters of the extended Vdn model, $B_\mathrm{slope}$ and $B_\mathrm{offset}$, to the median values obtained from the calibration performed with Flagship data (label: fixed calibration);
secondly, allowing $B_\mathrm{slope}$ and $B_\mathrm{offset}$ to vary in the parameter space described by a 2D Gaussian distribution centred on their median values and given by the calibration with the Flagship simulation (label: relaxed calibration).

The two adopted approaches are meant to demonstrate the impact of the calibration that will be performed in \Cref{subsec:Vdn_analysis} on the cosmological forecast. In this work the constraints on the parameters $B_\mathrm{slope}$ and $B_\mathrm{offset}$ are indeed limited to the statistical relevance of the number counts of voids identified by means of the Flagship galaxies. The case in which the cosmological forecasts are computed fixing $B_\mathrm{slope}$ and $B_\mathrm{offset}$ to their exact calibrated values represents therefore an optimistic evaluation of the results that we may obtain in the future thanks to the usage of larger mock catalogues, or by means of a fully theoretical modelling of the tracer bias inside cosmic voids (see \Cref{subsec:methodology}).

The cosmological model considered for the analysis is characterised by a primordial comoving curvature power spectrum amplitude fixed to the Flagship simulation value, $A_{\rm s} = 2.11 \times 10^{-9}$. We followed the strategy to fix this parameter in order to mimic the future application to real data, which will be supported by the impressive constraints obtained from the study of CMB anisotropies by \citet{Planck2018}. Thanks to this approach, for each MCMC step we derived $\sigma_8$, i.e. the root mean square mass fluctuation in spheres with radius $8 \ h^{-1} \ \mathrm{Mpc}$. We relied on \texttt{CAMB} to compute this quantity as a derived parameter, which depends on all the cosmological parameters involved in the evolution of the matter power spectrum $P_\mathrm{m}(k)$.

The density parameter $\Omega_{\rm m}$ is computed as the sum of cold dark matter, baryon and neutrino energy densities, ${\Omega_{\rm m}=\Omega_{\rm cdm}+\Omega_{\rm b}+\Omega_\nu}$, and its variation in the Bayesian statistical analysis is balanced by the changing of the DE density parameter, $\Omega_{\rm de}$, to keep flat the universe geometry, $\Omega_{\rm de}=1-\Omega_{\rm m}$.

The implementation of massive neutrinos in the MCMC analysis was performed considering the sum of the mass of neutrinos as a free parameter in the cosmological model. Neutrinos were modelled with one massive eigenstate and two massless ones, assuming an effective number of neutrino species $N_\mathrm{eff}=3.04$ \citep{Froustey2020,Bennett2020} and relating the neutrino mass to the neutrino density parameter as \citep{Mangano2005}:
\begin{equation}
    \Omega_\nu = \frac{M_\nu}{93.14 \ h^2 \ \mathrm{eV}} \, \text{,}
\end{equation}
where we denote $M_\nu = \sum m_\nu$ as the sum of the neutrino mass eigenstates. 

Since the thermal free-streaming of massive neutrinos suppresses density fluctuations, the abundance of voids changes with massive neutrinos, with respect to the massless neutrinos case \cite[see e.g.][for a discussion]{kreisch2018,schuster2019,Contarini2020}. 
We included the variation of the neutrino density parameter, $\Omega_\nu$, in the MCMC analysis, by keeping the value of the total matter density $\Omega_{\rm m}$ fixed (see \Cref{subsec:Flagship}), thus rescaling consistently the cold dark matter density parameter $\Omega_{\rm cdm}$. We tested the effect of considering both $\Omega_{\rm m}$ and $\Omega_\nu$ as free parameters of the model, finding a strong degeneracy between the two. We chose therefore to separate into different scenarios the models having either $\Omega_{\rm m}$ or $\Omega_\nu$ unconstrained: our goal is to investigate the sensitivity of the void size function to these two parameters independently, aiming at using this void statistic in combination with other cosmological probes.

We relied on \texttt{CAMB} for the computation of the total matter power spectrum used to predict the theoretical model of the void size function. The region of the parameter space characterised by a DE equation of state with $w_0+w_a>0$ is not covered by \texttt{CAMB}.

\section{Results}\label{sec:results}
The aim of this section is to compare our theoretical predictions with the void size function measured from the Flagship simulation. We then provide forecasts for the \Euclid survey, using a Bayesian statistical analysis to predict constraints on the parameters of the DE equation of state, modelling the void size function according to the theoretical prescriptions reported in \Cref{sec:theory_&_methods}.

\subsection{Void size function analysis}\label{subsec:Vdn_analysis}

To compare the theoretical void size function with the number counts of voids measured in the galaxy distribution, we need to convert the threshold $\delta_\mathrm{v,tr}^\mathrm{NL}$ fixed in measurements to the corresponding one in the matter distribution, as described in \Cref{subsec:methodology}.
First of all, we verified the calibration of the relation $\mathcal{F}(b_\mathrm{eff})$ reported in \Cref{b_punct_formula} using the Flagship simulation. To this end, we extracted the value of $B_\mathrm{slope}$ and $B_\mathrm{offset}$ by leaving them as free parameters with uniform priors of the extended Vdn model and fitting the measured void number counts in the selected redshift bins, considering also a Gaussian prior for $b_\mathrm{eff}$ at each redshift. We notice that, since the error on the effective bias only corresponds to a few percent of its value, the variation allowed for this parameter during the fit is small. All the remaining cosmological parameters were kept fixed to the Flagship simulation values during this calibration.

With this prescription we obtained the confidence levels reported on the left panel of \Cref{fig:bias_relation}, for the void size function measured in both real and redshift space in light blue and orange, respectively. The resulting coefficients for the calibrated relations are:
\begin{gather}\label{eq:calibration}
\mathcal{F}(b_\mathrm{eff}) = (0.96 \pm 0.04)\  b_\mathrm{eff} + (0.44 \pm 0.07) \ \text{ and} \\
\mathcal{F}(b_\mathrm{eff}) = (0.96 \pm 0.03) \ b_\mathrm{eff} + (0.26 \pm 0.06) \ \text{ ,} 
\end{gather}
for the redshift-space and the real-space void abundance, respectively.

We show on the right panel of \Cref{fig:bias_relation} the corresponding linear relations obtained with these calibrations, with a shaded area representing an uncertainty of $2 \sigma$. As a comparison, we present in the same plot the values computed for $b_\mathrm{punct}$, leaving it as the only free parameter of the model and fitting separately the measures at different redshifts.
This analysis is aimed at testing the precision of the calibrated relations for each redshift: in the right plot of \Cref{fig:bias_relation} the markers with best match to the linear relations correspond in \Cref{fig:void_size_functions} to the redshift bins for which the calibrated model more accurately reproduces the measured void number counts, while points that depart from the linear relationship in \Cref{fig:bias_relation} (right plot) will lead to a slightly worse agreement between theory and model in \Cref{fig:void_size_functions}.

Finally, we report also the calibration performed in \citep{contarini2019} using the dark matter haloes of the COupled Dark Energy Cosmological Simulations \citep[CoDECS,][]{baldi2012}, and represented in grey in the right panel of \Cref{fig:bias_relation}.
At lower redshifts the calibration we measure in this paper is in good agreement with the calibration from the CoDECS simulation, characterised by a cosmology consistent with the results of the seven-year Wilkinson Microwave Anisotropy Probe \citep[WMAP7,][]{Komatsu2011}, but it slightly deviates from the latter at higher redshift values. The reason for this minor deviation is twofold. Firstly it is linked to the kind of cosmic tracers (i.e. dark matter haloes or galaxies) and the selection criteria (i.e. minimum mass or magnitude) used to identify voids \citep{Contarini2020}. Secondly it is related to the fact that in \citet{contarini2019} the calibration was performed for redshift from $0$ to $1$, while here we are testing this relationship beyond this range. The physics underlying the function $\mathcal{F}(b_\mathrm{eff})$ and its relation with the mass tracers used to identify voids will be investigated in future papers.

More importantly, since the void size function will be measured on real data from the \Euclid survey, we have to deal with voids detected in redshift space. The overall effect of RSD on voids, relevant for the void size function, is an apparent enlargement of the voids' volume, due to the elongation along the line of sight. This is reflected in a mean shift of the measured void size function towards greater radii.
Even if this effect can in principle be theoretically modelled \citep{pisani2015b,Correa2020},  
we decide to parametrise it empirically as described below.
Indeed, the theoretical approach requires knowledge of the void matter density profile for the entire void population, which has to be characterised in simulations and may introduce some model dependencies. We found that the parametrisation of $\mathcal{F}(b_\mathrm{eff})$ can be exploited to encapsulate also the modifications on the void sizes caused by the enlargement of cosmic voids in redshift space. This approach has the advantage of being both simple to model and robust, allowing us to take into account, with the same parameter, both the impact of tracer bias in voids and of the RSD. Moreover, this approach is fully agnostic and does not require any assumption about the void density profile, nor any other modelling, making it particularly suited to survey analyses.

It is worth noting that the relation obtained for voids in redshift space shows a greater offset but almost the same slope with respect to its analog in real space. This difference reflects the increase of void sizes in redshift space. It also opens the way to test theoretical implementations in future work, indicating that a simple modelling of those effects should suffice to extract robust constraints.

Equipped with these calibrated relations, we now have all the elements necessary to compare the measured void size function with the theoretical predictions given by the extended Vdn model, in which the underdensity threshold is converted as described in \Cref{subsec:methodology}. \cref{fig:void_size_functions}
provides the main results of our Flagship analysis. We show the comparison between the measured void number counts and the corresponding theoretical void size functions, both in real and redshift space, for the six equi-populated bins in redshift. The Poissonian errors related to the data are represented by the error bars, while the uncertainty related to the theoretical model is shown as a shaded region. The latter is computed associating an error to $\mathcal{F}(b_\mathrm{eff})$ given by the interval delimited by the coloured bands in \Cref{fig:bias_relation}. The residuals are reported at the bottom of each sub-plot and are calculated as the difference from the theoretical model, in units of the data errors. The latter show an excellent agreement between simulated data and theoretical models, even when considering voids identified in the Flagship galaxy catalogue in redshift space. The measured void number counts are indeed within an uncertainty of $2\sigma$, shown by the hatched coloured bands in the bottom panels, represented in units of the data errors. 

To test the goodness of the fits shown in \Cref{fig:void_size_functions} we computed the reduced $\rchi^2$ using the weighted sum of squared deviations of the two data sets from their corresponding models and dividing the results by the degrees of freedom of the two systems, $\nu = N_{\rm bin} - N_{\rm par}$, where $N_{\rm bin}$ is the number of bins in radius used to compute the void counts and $N_{\rm bin}$ is the number of free parameters of the model. In our case $N_{\rm bin}=50$ and $N_{\rm par}=2$, since we are fitting the void counts simultaneously for all the redshift shells and we are considering $B_\mathrm{slope}$ and $B_\mathrm{offset}$ only as free parameters of the model. The results are $\rchi_\nu^2=1.60$ and $\rchi_\nu^2=1.02$ for real and redshift space, respectively.

\begin{figure*}
\centering
\includegraphics[width=0.46\textwidth]{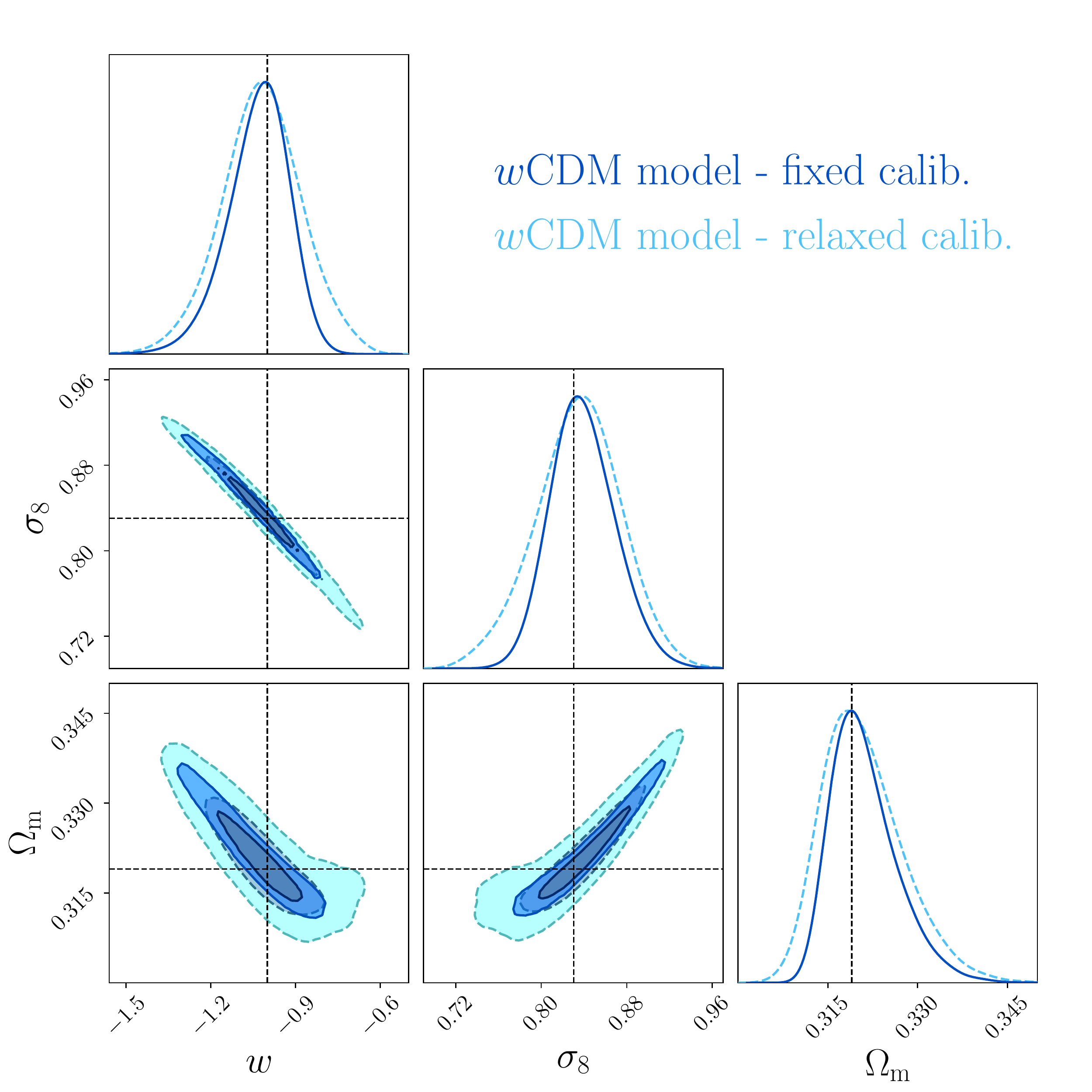} 
\hspace{0.5cm}
\includegraphics[width=0.46\textwidth]{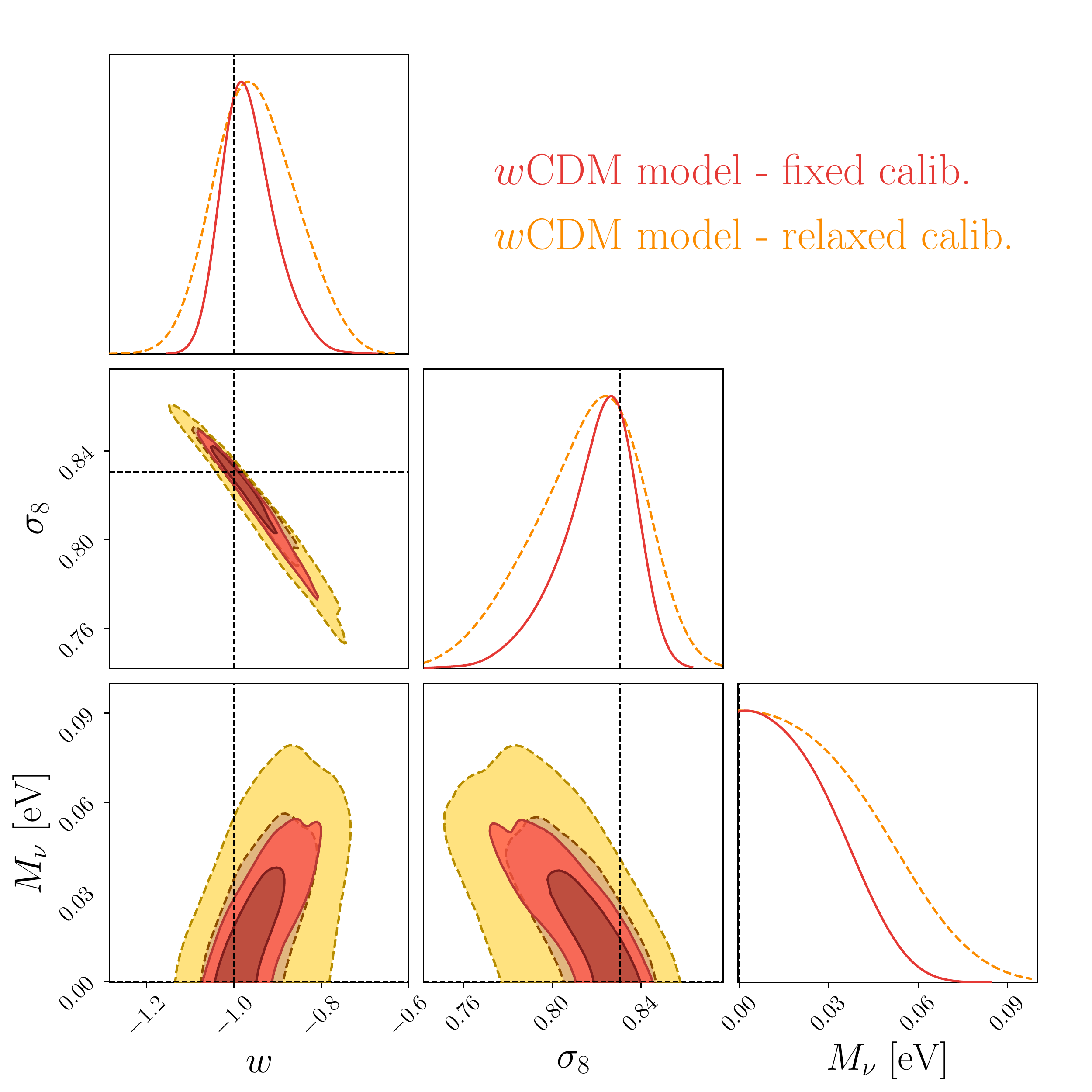}
\caption{Cosmological forecasts for the \Euclid mission from the void size function for the $ w$CDM model, characterised by a DE component described by a constant $w$. The contours represent the $68\%$ and $95\%$ confidence levels obtained by means of the Bayesian statistical analysis described in \Cref{subsec:forecasts_theory}.
\textit{Left}: forecasts for a cosmological model with $w$ and $\Omega_{\rm m}$ as free cosmological parameters. We report the constraints obtained by fixing the calibration parameters with blue contours marked by a solid line and the results obtained by relaxing the calibration constraints with light-blue contours marked by a dashed line (see \Cref{subsec:cosmological_models}).
\textit{Right}: forecasts for a cosmological model with $w$ and $M_\nu$ as free cosmological parameters. We represent the results of the fixed calibration case as red confidence contours having solid borders and those of the relaxed calibration case as orange contours having dashed borders.
For each plot we show also the constraints on $\sigma_8$, computed as a derived parameter.
The true values of the parameters are shown by a black dashed line.}
\label{fig:final_constraints1}
\end{figure*}

\begin{figure*}
\centering
\includegraphics[width=0.49\textwidth]{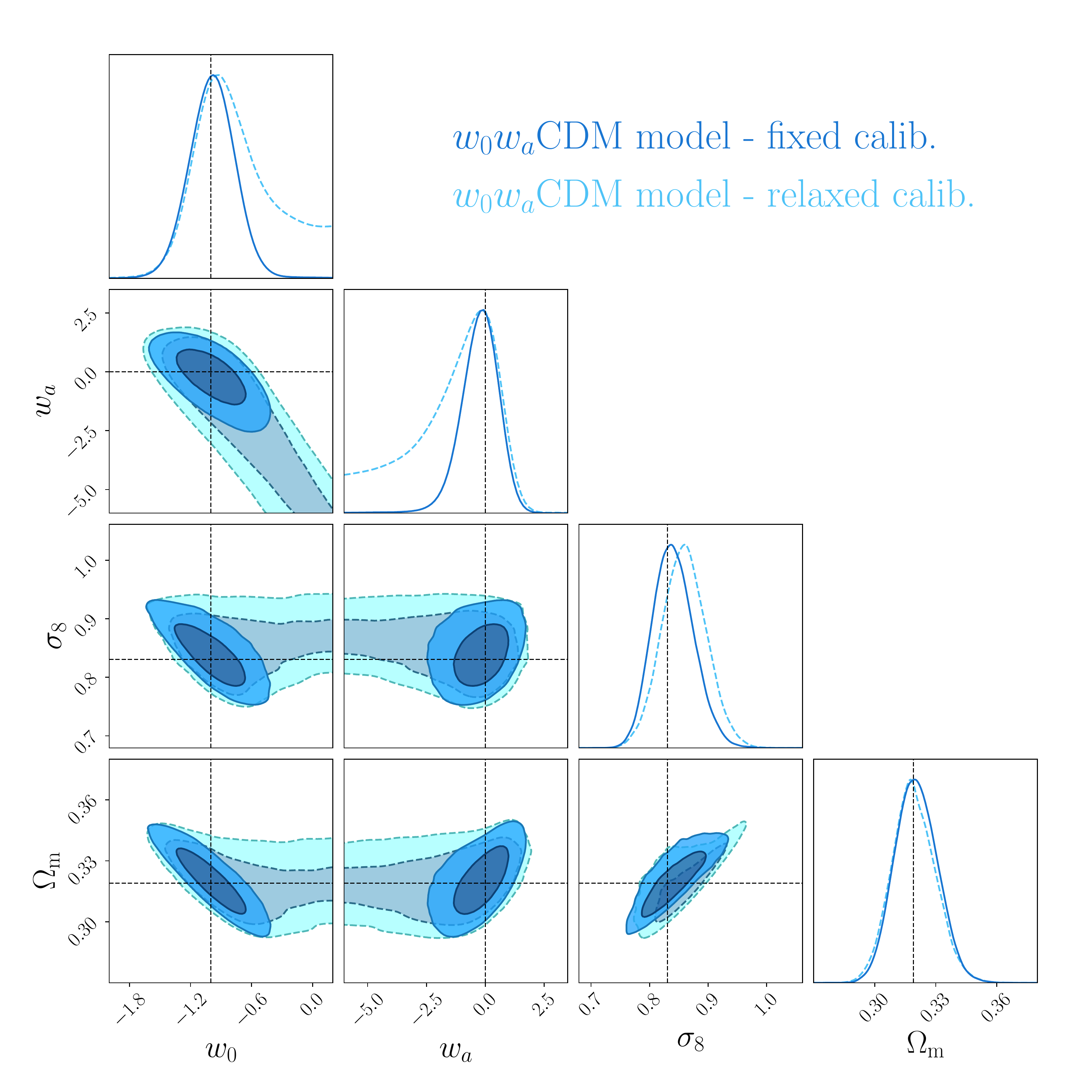} 
\hfill
\includegraphics[width=0.49\textwidth]{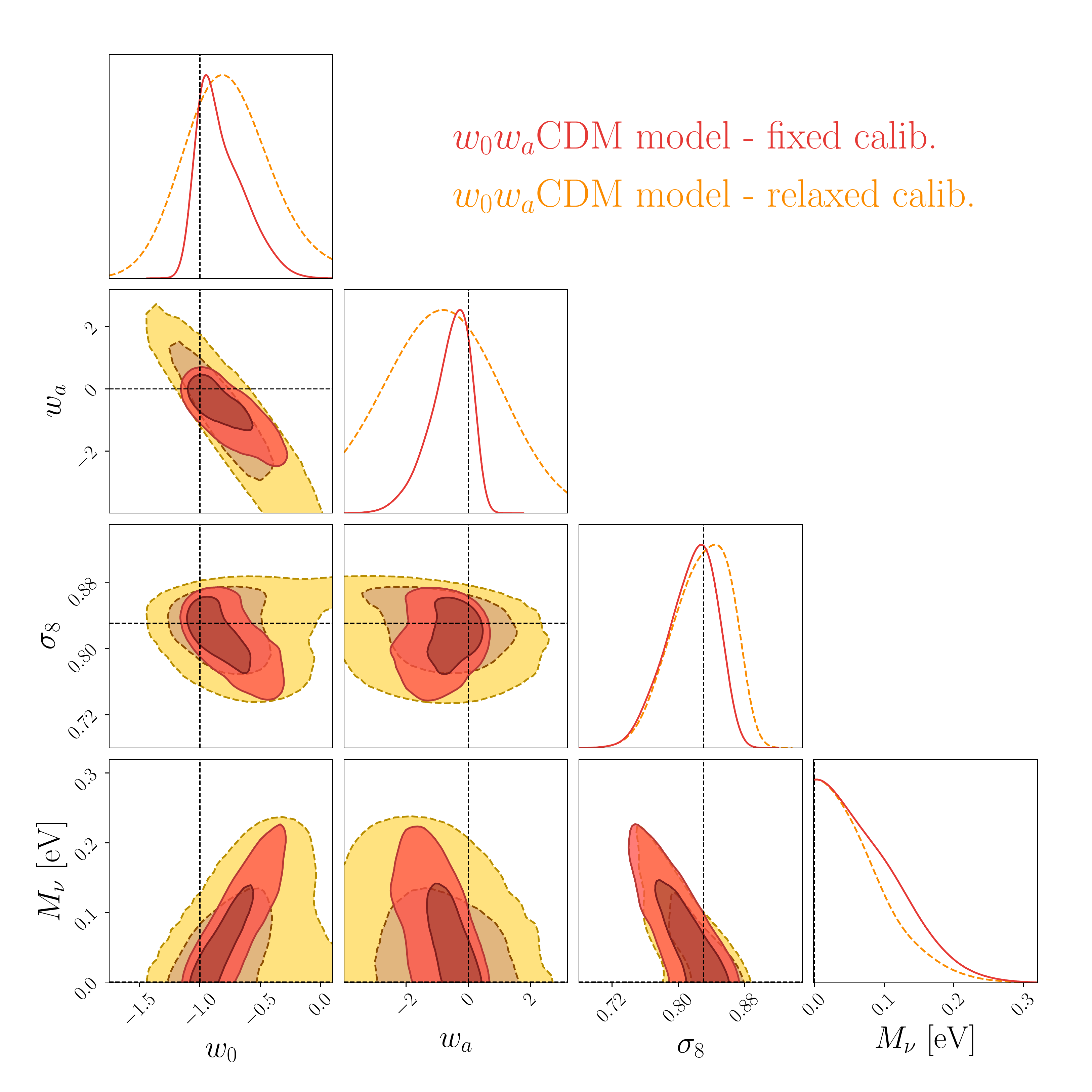}
\caption{Same as \Cref{fig:final_constraints1} but for the cosmological model labelled as $ w_0 w_a$CDM, having a dynamical DE component described by the CPL parametrisation (see \Cref{subsec:cosmological_models}).}
\label{fig:final_constraints2}
\end{figure*}

\subsection{Cosmological forecasts}\label{subsec:cosmological_forecats}

In this section we provide the cosmological forecasts obtained using the void size function in redshift space in the perspective of the \Euclid mission. We applied the statistical analysis described in \Cref{subsec:forecasts_theory} to derive constraints on the parameters of the two cosmological models analysed, labelled as $w$CDM and $ w_0 w_a$CDM, following the two approaches described in \Cref{subsec:cosmological_models}.
For the model $w$CDM we assumed a flat prior for all the remaining free cosmological parameters of the model, and for the model $ w_0 w_a$CDM we assumed a Gaussian prior distribution with standard deviation $\sigma=5$ for $w_0$ and $\sigma=15$ for $w_a$, both centred on the true values of these parameters, given by the Flagship simulation cosmology ($w_0=-1$, $w_a=0$). We preferred to use very wide Gaussian priors instead of uniform ones to improve the numerical stability of the whole pipeline, but we tested that uniform priors yield consistent results. The remaining cosmological parameters analysed in this work ($\Omega_{\rm m}$ and $M_\nu$) were included in the void size function modelling with uniform prior distributions.

In \Cref{fig:final_constraints1} we present the $68\%$ and $95\%$ confidence levels of the constraints on the model $w$CDM. In the left plot we show the \Euclid forecasts from a void size function model characterised by $w$ and $\Omega_{\rm m}$ as free cosmological parameters. We represent with different colours and borders the results obtained with the two approaches described in \Cref{subsec:cosmological_models}: in blue with solid contours the forecasts obtained by fixing the extended Vdn parameters $B_\mathrm{slope}$ and $B_\mathrm{offset}$, in light-blue with dashed contours those obtained by relaxing the calibration constraints by means of a 2D Gaussian prior on $B_\mathrm{slope}$ and $B_\mathrm{offset}$, which distribution is represented in the left panel of \Cref{fig:bias_relation}.
In the right plot we represent the same forecasts but considering a void size function model with the neutrino total mass $M_\nu$ as free parameter instead of the matter density $\Omega_{\rm m}$. In this case we show the fixed and the relaxed calibration approach results in red and orange, respectively.
In both the presented cases $\sigma_8$ is computed as derived parameter. As expected, the effect of relaxing the calibration constraints is to broaden the confidence contours.

\begin{table*}
    \caption{
    Cosmological forecasts computed for the \Euclid mission from the void size function for the cosmological model $w$CDM. In this table we report the results of the two analysis strategies adopted in this work: considering the parameters $B_\mathrm{slope}$ and $B_\mathrm{offset}$ fixed to the respective median calibrated values (label: fixed calib.) or with a multivariate Gaussian with the same median value but a constraining power given by the calibration procedure with Flagship (label: relaxed calib.).
    For each of the two cases we present, in the upper and lower line, the forecasts obtained fixing $M_\nu$ or $\Omega_{\rm m}$ to the Flagship simulation true values, respectively.
    All the constraints are reported with errors with a $1\sigma$ confidence level.
    }
    \centering
        \begin{tabular}{ccccccc}
        \toprule
        \noalign{\vspace{0.075cm}}
		Model & $w$ & $\sigma_8$ & $\Omega_{\rm m}$ & $M_\nu \, [\mathrm{eV}]$ & $B_\mathrm{slope}$ & $B_\mathrm{offset}$ \\ 
		\noalign{\vspace{0.1cm}}
		\toprule
		\noalign{\vspace{0.15cm}}
        \multirow{3}{*}{\small{fixed calib.}} & $-1.01^{+0.09}_{-0.11}$ & $0.83 \pm 0.03$ & $0.319^{+0.005}_{-0.004}$ & $0$ & $0.96$ & $0.44$ \\ 
        \noalign{\vspace{0.2cm}}
		\cline{2-7}
		\noalign{\vspace{0.2cm}}
		& $-0.99^{+0.06}_{-0.04}$ & $0.83^{+0.1}_{-0.2}$ & $0.319$ & $ <0.03 $ & $0.96$ & $0.44$ \\ 
		\noalign{\vspace{0.2cm}}
		\cline{1-7}
		\noalign{\vspace{0.2cm}}
		\multirow{3}{*}{\small{relaxed calib.}} & $-1.0 \pm 0.1$ & $0.84 \pm 0.04$ & $0.318^{+0.008}_{-0.005}$ & $0$ & $0.96 \pm 0.02$ & $0.44 \pm 0.04$ \\ 
		\noalign{\vspace{0.2cm}}
		\cline{2-7}
		\noalign{\vspace{0.2cm}}
		& $-0.98^{+0.10}_{-0.07}$ & $0.83^{+0.02}_{-0.03}$ & $0.319$ & $ <0.06 $ & $0.95 \pm 0.02$ & $0.46 \pm 0.04$ \\  
		\noalign{\vspace{0.15cm}}
		\hline
		\bottomrule
		\end{tabular}
    \label{tab:cosm_constraints}
\end{table*}

\begin{table*}
    \caption{
    Same as \Cref{tab:cosm_constraints} but for the $w_0w_a$CDM scenario. In this case we present in the last column also the values computed with \Cref{eq:FoM} to estimate the FoM for the DE equation of state.}
    \centering
        \begin{tabular}{ccccccccc}
        \toprule
        \noalign{\vspace{0.075cm}}
		Model & $w_0$ & $w_a$ & $\sigma_8$ & $\Omega_{\rm m}$ & $M_\nu \, [\mathrm{eV}]$ & $B_\mathrm{slope}$ & $B_\mathrm{offset}$ & $\mathrm{FoM}_{w_0,w_a}$\\ 
		\noalign{\vspace{0.1cm}}
		\toprule
		\noalign{\vspace{0.15cm}}
		\multirow{3}{*}{\small{fixed calib.}} & $-1.0 \pm 0.2$ & $-0.1^{+0.7}_{-0.9}$ & $0.84^{+0.04}_{-0.03}$ & $0.32 \pm 0.01$ & $0$ & $0.96$ & $0.44$ & $4.9$ \\  
		\noalign{\vspace{0.2cm}}
		\cline{2-9}
		\noalign{\vspace{0.2cm}}
		& $-1.0^{+0.2}_{-0.6}$ & $-0.1^{+0.3}_{-0.8}$ & $0.83^{+0.02}_{-0.03}$ & $0.319$ & $<0.08$ & $0.96$ & $0.44$ & $17$ \\  
		\noalign{\vspace{0.2cm}}
		\cline{1-9}
		\noalign{\vspace{0.2cm}}
		\multirow{3}{*}{\small{relaxed calib.}} & $-0.8^{+1.6}_{-0.6}$ & $-0.9^{+3.6}_{-9.6}$ & $0.86 \pm 0.04$ & $0.32 \pm 0.01$ & $0$ & $ 1.01^{+0.03}_{-0.04}$ & $0.35^{+0.08}_{-0.05}$ & $0.78$ \\   
		\noalign{\vspace{0.2cm}}
		\cline{2-9}
		\noalign{\vspace{0.2cm}}
		& $-0.9^{+0.3}_{-0.2}$ & $-0.5^{+0.9}_{-1.3}$ & $0.86^{+0.02}_{-0.05}$ & $0.319$ & $<0.08$ & $0.99^{+0.01}_{-0.04}$ & $0.38^{+0.07}_{-0.01}$ & $2.3$ \\ 
		\noalign{\vspace{0.15cm}}
		\hline
		\bottomrule
		\end{tabular}
    \label{tab:cosm_constraints_FoM}
\end{table*}

In \Cref{fig:final_constraints2} we show the same contours represented in \Cref{fig:final_constraints1} but considering the $w_0w_a$CDM scenario. The free cosmological parameters of the void size function model are the coefficients of the DE equation of state, $w_0$ and $w_a$, together with $\Omega_{\rm m}$ (left plot) or $M_\nu$ (right plot). Also in this case the relaxation of the constraining condition of the calibration parameters causes an enlargement of the confidence contours. In this scenario however, the strongest impact of the calibration constraints is on the $w_0$--$w_a$ parameter plane, in particular along the diagonal where these parameters become degenerate. The effect of the calibration constraints on $\Omega_{\rm m}$ and $M_\nu$ has a lower impact.

In \Cref{tab:cosm_constraints,tab:cosm_constraints_FoM} we report the values, with relative $1\sigma$ errors, of the cosmological constraints derived for the $w$CDM and $w_0w_a$CDM scenario, respectively. The constraints on the sum of neutrino masses $M_\nu$ are expressed as a $1\sigma$ upper limit. For each table we show the results for the two approaches followed in this paper: fixing and relaxing the calibration constraints on the void size function model.
The calibration parameter are reported in the columns $B_\mathrm{slope}$ and $B_\mathrm{offset}$ for completeness. 
Notice that each quantity reported without any uncertainty is considered fixed in the specific scenario presented in that table row. 

For the $w_0w_a$CDM scenario, in order to evaluate the constraining power of the void size function on the DE equation of state, we derived the figure of merit (FoM) for the coefficients of the CPL parametrisation $w_0$ and $w_a$. 
We computed this value by following \cite{Wang2008}:
\begin{equation}\label{eq:FoM}
\mathrm{FoM}_{w_0,w_a} = \frac{1}{\sqrt{\mathrm{det} \ \mathrm{Cov}(w_0, w_a) }} \, \text{,}
\end{equation}
where $\mathrm{Cov}(w_0, w_a)$ represents the covariance matrix of the DE equation of state parameters. We notice that this definition is in agreement with the one adopted in \cite{EC2020}. We underline that, contrary to other constraint accuracy indicators (e.g. the relative or percentage errors), the higher the FoM value, the better the precision on the analysed cosmological parameters. We report this quantity in the last column of \Cref{tab:cosm_constraints_FoM}.

\begin{figure*}
\centering
\includegraphics[width=0.47\textwidth]{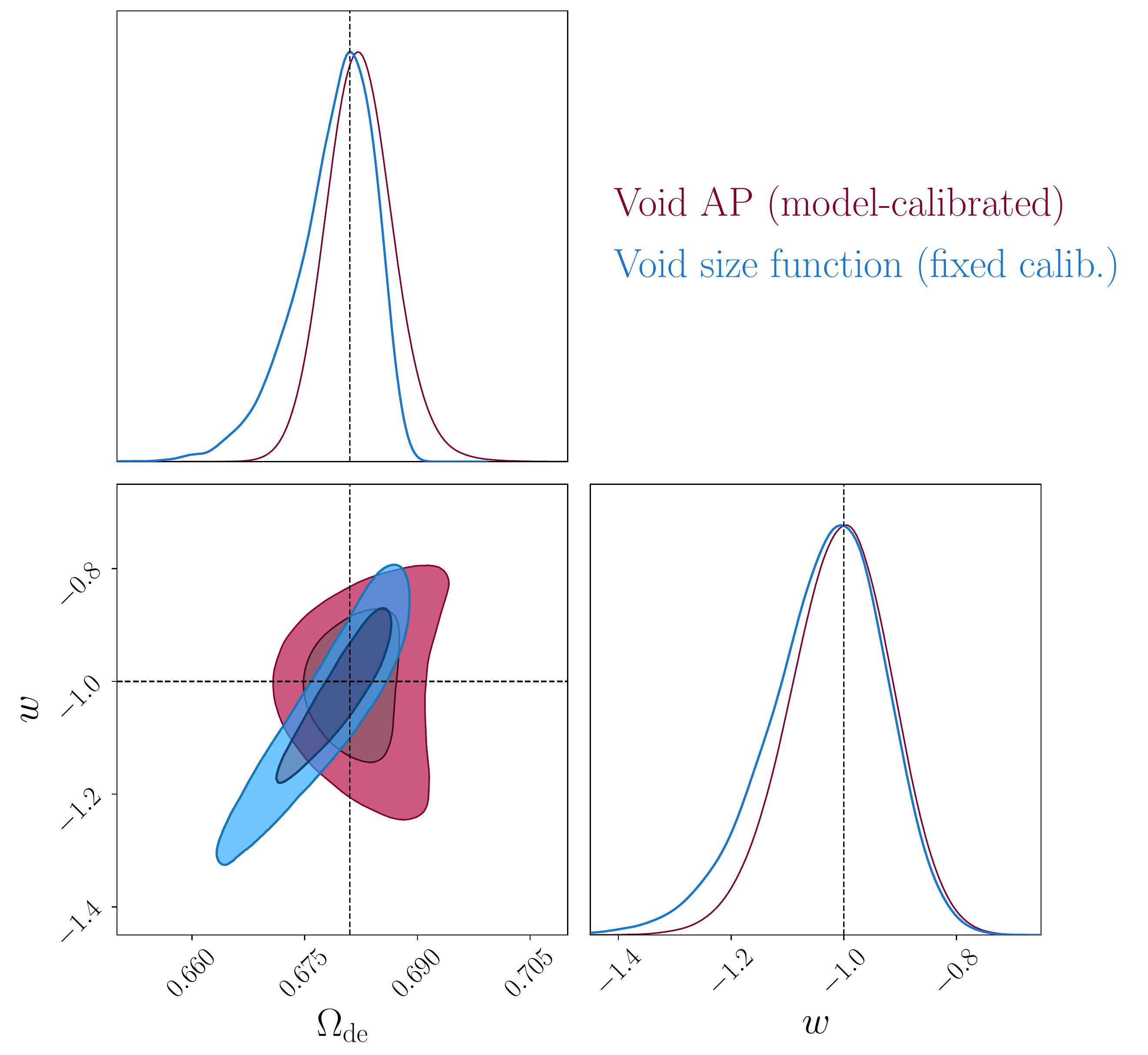} 
\hspace{0.5cm}
\includegraphics[width=0.46\textwidth]{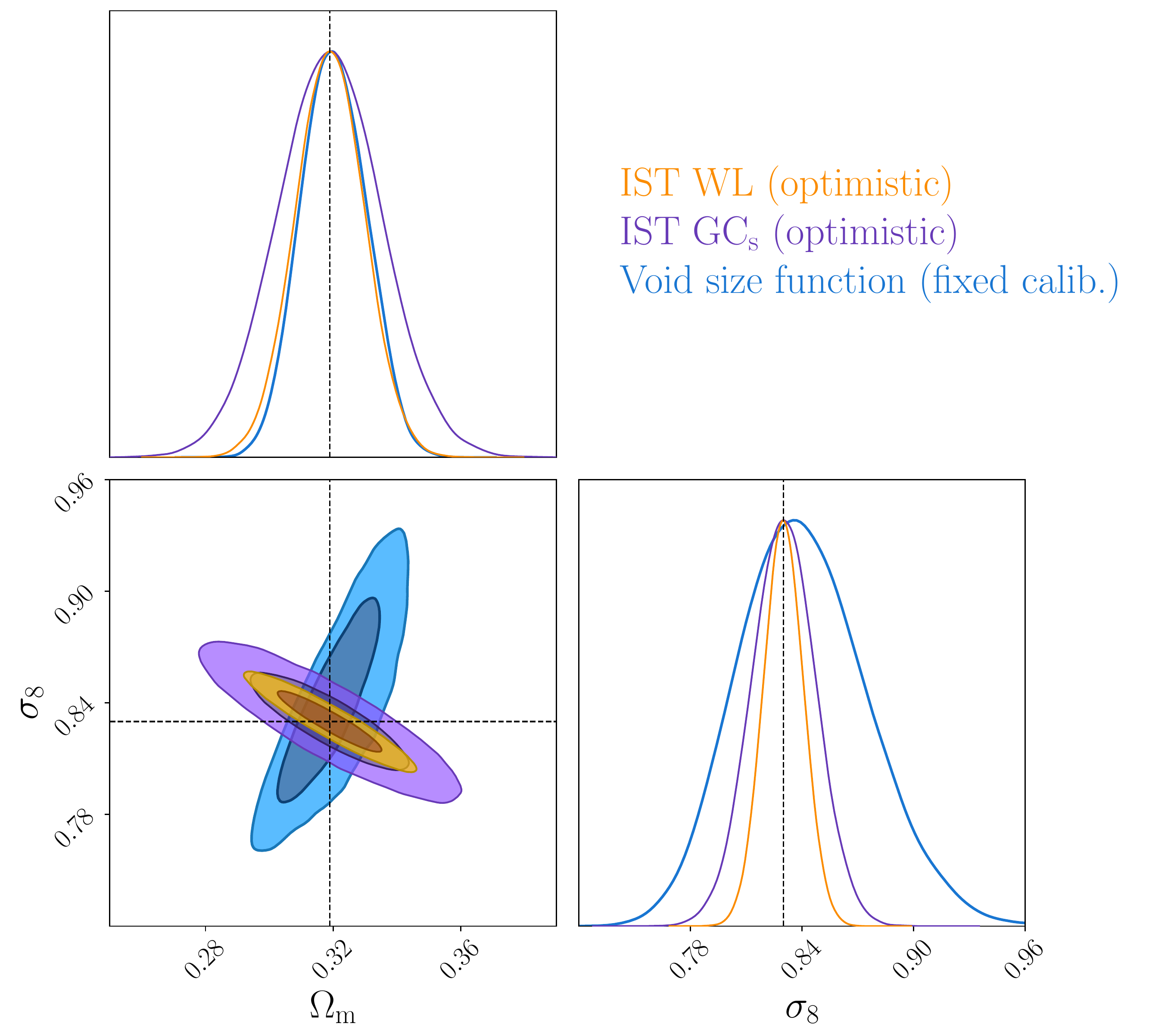}
\caption{Comparison between the 68\% and 95\% confidence levels computed in this work with the void size function and different \Euclid forecasts. \textit{Left}: cosmological constraints on the $\Omega_{\rm de}$--$w$ plane computed in this work (in blue) considering a $w$CDM scenario with fixed calibration parameters and in \citet{EuclidHamaus2021} (in magenta), modelling the void-galaxy cross-correlation function in redshift space, with a model-calibrated approach. \textit{Right}: cosmological constraints on the $\Omega_{\rm m}$--$\sigma_8$ plane computed in this work considering a $w_0w_a$CDM scenario (in blue) with fixed calibration parameters and the marginalised IST Fisher forecasts computed in the optimistic setting with spectroscopic galaxy clustering (in purple) and weak lensing (in orange).}
\label{fig:forecasts_comparison}
\end{figure*}

As a first exploration of the cosmic void statistics combined power, we now compare the forecasts from the void size function provided in this work with other \Euclid forecasts. We present as a first comparison the results of the $\Omega_{\rm de}$--$w$ confidence contour with the model-calibrated forecasts presented in \citet{EuclidHamaus2021}. The latter are computed by modelling the observable distortions of average shapes in redshift space via RSD and the Alcock--Paczy{\'n}ski effect, for voids to be measured in the \Euclid spectroscopic galaxy distribution. Contrary to the model-independent case, in the presented approach the nuisance parameters of the model have been calibrated by means of Flagship data. In this comparison we consider the $w$CDM scenario with fixed neutrino mass and we focus on the $\Omega_{\rm de}$--$w$ parameter space. Given the assumption of flat spatial geometry, to compute the corresponding $\Omega_{\rm de}$ forecasts, we converted the $\Omega_{\rm m}$ obtained in the MCMC analysis as $\Omega_{\rm de}=1-\Omega_{\rm m}$.

As a second comparison we take the results of Fisher analysis reported by the inter-science taskforce for forecasting \citep[IST:F,][]{EC2020} obtained in the optimistic setting for the single probes weak lensing and galaxy clustering. We considered in this case the $w_0w_a$CDM scenario with fixed neutrino mass and we focus on the $\Omega_{\rm m}$--$\sigma_8$ degeneracy. To compute the IST confidence contour we made use of the publicly available\footnote{See \url{https://github.com/euclidist-forecasting/fisher_for_public}.} Fisher matrices and we marginalised on the parameters not reported in the plot with the code \texttt{CosmicFish} \citep{CosmicFish2016}. We recall that amplitude of density fluctuations at $z=0$, $\sigma_8$, is computed as a derived parameter in our analysis and its variation is given by the modifications caused by the free cosmological parameters of the model to the total matter power spectrum. We also stress the fact that a larger set of cosmological parameters is used in IST forecasts. This includes in particular the baryon matter energy density, $\Omega_{\rm b}$, the dimensionless Hubble parameter, $h$, and the spectral index of the primordial density power spectrum, $n_\mathrm{s}$. The impact on forecasts when including these parameters in the model will be tested in future work.

We show the presented comparisons in \Cref{fig:forecasts_comparison}, representing in blue the forecasts obtained in this work considering a void size function model with fixed calibrated parameters. In the left panel we compare our results with the $\Omega_{\rm de}$--$w$ confidence contour computed with the model-calibrated forecasts presented \citet{EuclidHamaus2021} (in magenta). In the right panel we show instead the comparison of $\Omega_{\rm m}$--$\sigma_8$ confidence contour provided by IST forecasts considering the optimistic setting for weak lensing (in orange) and galaxy clustering (in purple). See \Cref{AppendixB} for an analogous comparison considering the cosmological forecasts presented above but with less optimistic modelling approaches.

In both panels we can appreciate the comparable extension of the presented contours and in the latter we can notice in particular the strong complementarity of the void size function forecasts with those of the \Euclid standard probes.
While a more accurate analysis would require proper accounting of covariance between analysed cosmological constraints, \Cref{fig:forecasts_comparison} shows how the presented probes explore the parameter space differently and motivates investigation on combination to be performed in future works.

\section{Conclusions and discussion}\label{sec:conclusions}

In this work we presented state-of-the-art forecasts for cosmological constraints from the void size function to be expected from the \Euclid mission. We measured the void number counts from the Flagship mock galaxy spectroscopic catalogue in redshift bins and matched the measurements with the theoretical definition given by the Vdn model \citep{SVdW2004,jennings2013}. We employed an extension of the Vdn model that conservatively accounts for the effects of the galaxy large-scale bias, $b_\mathrm{eff}$, on the void effective radii. With this method, we parametrised the Vdn model's characteristic threshold $\delta_\mathrm{v}^\mathrm{L}$ according to the prescriptions of \citet{contarini2019}, also verifying the calibration of the function $\mathcal{F}(b_\mathrm{eff})$. The parametrisation method further allowed us to account for the modifications on the void sizes caused by the volume change of cosmic voids in redshift space. 

We showed that the extended Vdn calibrated on Flagship data is effective in predicting the measured void number counts both in real and redshift space. Indeed we obtained a remarkable agreement between the measured and predicted void size functions, for all the redshift bins and all the spatial scales considered in our analysis. We also performed a MCMC analysis, estimating the constraints from void number counts on two main cosmological models: assuming in one case a scenario characterised by a constant equation-of-state parameter ($w$CDM) and in the other case a scenario with a dynamical DE component described by the CPL parametrisation ($w_0w_a$CDM). For each scenario we presented the \Euclid cosmological forecasts considering both approaches: by fixing the extended Vdn model parameters and by relaxing their boundaries to those provided by the calibration with Flagship mock catalogues. The former represents the ideal situation in which the simulations used to calibrate the void size function model allow us to have no uncertainties nor systematic errors on the calibration parameters; or alternatively, the case in which the value of the tracer bias inside voids is fully determined thanks to theoretical modelling.

In the $w$CDM scenario we forecasted relative percentage errors on the constant DE component, $w$, below the $10\%$ for each analysed case. In the $w_0w_a$CDM scenario, with the optimistic approach of fixing the model calibration parameters, we computed a $\mathrm{FoM}_{w_0,w_a}$ equal to $4.9$ or $17$, in the case of leaving $\Omega_{\rm m}$ or $M_\nu$, respectively, as additional free cosmological parameters of the model.
As a reference, the corresponding FoM values computed by the IST:F for spectroscopic galaxy clustering and weak lensing are $55$ and $44$, respectively (see Table 13 in \citealp{EC2020}, for flat-$w_0w_a$CDM cosmology and optimistic scenario).

The marginalised constraints on the derived parameter $\sigma_8$ are lower than $5\%$ in every analysed case, while the relative errors on $\Omega_{\rm m}$ are of the order of $2\%$ in the $w$CDM scenario and of $3\%$ in the $w_0w_a$CDM scenario. The $1\sigma$ upper limit on $M_\nu$ is instead of $0.03 \ \mathrm{eV}$ in the most optimistic case of the $w$CDM scenario and of $0.08 \ \mathrm{eV}$ in the $w_0w_a$CDM scenario. We recall that, in the cosmological models with free neutrino mass, the total matter energy density was fixed to the Flagship simulation true value, therefore the degeneracy of $\Omega_\nu$ with $\Omega_{\rm m}$ is not considered in the results.

Our analysis showcases the impressive constraining power of the void size function from the \Euclid survey, strongly complementing the \Euclid primary probes. This complementarity will make the combination powerful in particular for weak lensing and galaxy clustering, additionally enhancing robustness to systematic effects in both cases.

In this work we considered extremely conservative assumptions when analysing the void sample.
Such conservative assumptions dramatically reduce the statistical power of our void catalogues, to ensure strong reliability: in the future, modelling improvements will allow a more efficient void selection, critically enhancing results while maintaining full robustness. 

Among the conservative choices in modelling the void size function and in building the likelihood we recall the treatment of both the threshold value and of the minimum void radius accepted for the analysis.
In particular, we selected a low value of the underdensity threshold ($\delta_\mathrm{v,tr}^\mathrm{NL}=-0.7)$ to avoid the shallowest voids in the sample, characterised by higher Poissonian noise contamination. We also strictly restricted the range of considered radii to avoid modelling poorly sampled voids of the Flagship galaxy catalogue, in order to prevent the inclusion of spatial scales affected by a loss of void counts. This conservative approach allows us to have a sample of voids composed by a limited number of objects but characterised by high purity. Different techniques will be tested in the future to 
better model the scales affected by numerical incompleteness \citep[see e.g.][]{cousinou2019} and include them in the analysis, safely obtaining access to much larger statistics. A better modelling of these effects will lead to further improvements in the constraining power of the void size function.

Further future prospects to expand this work include exploiting void number count forecasts to predict constraints from other void applications (such as the stacked void-galaxy cross-correlation function, see \citealp[]{EuclidHamaus2021}, void lensing, see \citealp[]{Bonici2022}, the void-void correlation function, see \citealp[]{Kreisch2021}), and subsequently combine joint constraints from voids with other \Euclid probes (primary and not, e.g. galaxy clustering, galaxy weak lensing, cluster counts and clustering, baryonic
acoustic oscillations, supernova distance measurements, CMB cross-correlations, etc.).

Other areas to explore include considering other cosmological parameters for the likelihood modelling, a more realistic treatment of observational effects (a more complex survey mask, a more realistic $\sigma_z$ and further survey-related systematic effects). 
Moreover, to prepare the application to real-survey data, the large-scale effective bias can be recovered from redshift-space catalogues. To this purpose, a number of methods are available (see e.g. \citealp[]{Chan2012, Sheth2013, Lazeyras2016, Sanchez2017b} and \citealp[]{desjacques_bias2016} for an extensive review), which would allow us to reliably model linear bias as a function of redshift, $b(z)$.
These methods can be exploited in the context of the \Euclid mission, deriving tight constraints on $b(z)$, especially from the cross-correlation between galaxy clustering and weak lensing, capable of breaking the degeneracies with cosmological parameters \citep{Tutusaus2020}. 

Finally, a purely theoretical treatment of the impact of RSD and of the Alcock--Paczy{\'n}ski effect on the void size function, based on work from companion papers \citep{EuclidHamaus2021}, as well as a comprehensive analysis of the systematic effects linked to these corrections is relevant to consider in future works.

For the future, a possible approach to model the measured \Euclid void number counts would be to apply directly the relation $\mathcal{F}(b_\mathrm{eff})$ calibrated on the redshift-space void sample of the Flagship mock (see Eq.~\ref{eq:calibration}). A close match of the galaxy properties in the mock catalogue with galaxies to be observed by \Euclid would make this approach particularly effective. Future works will aim to explore these methodologies and their impact on void constraints, along with testing the range of applicability of the $\mathcal{F}(b_\mathrm{eff})$ relation and its possible cosmological dependency (expected however to be mild from \citealp{Contarini2020}).

This paper -- with a first analysis on a full mock, the \Euclid Flagship simulation -- shows the impressive constraining capability of void number counts to tackle the properties of DE and neutrinos, demonstrating for the first time the feasibility of the technique with an end-to-end data-like application, and setting the ground for a robust use of the void size function for cosmology with \Euclid.  

\begin{acknowledgements}
We acknowledge the grant ASI n.2018-23-HH.0. SC, FM and LM acknowledge the use of computational resources from the parallel computing cluster of the Open Physics Hub (\url{https://site.unibo.it/openphysicshub/en}) at the Physics and Astronomy Department in Bologna.
GV is supported by Università degli Studi di Padova and in part by the project “Combining Cosmic Microwave Background and Large Scale Structure data: an Integrated Approach for Addressing Fundamental Questions in Cosmology”, funded by the MIUR Progetti di Rilevante Interesse Nazionale (PRIN) Bando 2017 - grant 2017YJYZAH.
AP is supported by NASA ROSES grant 12-EUCLID12-0004, and NASA grant 15-WFIRST15-0008 to the Nancy Grace Roman Space Telescope Science Investigation Team ``Cosmology with the High Latitude Survey''. NH is supported by the Excellence Cluster ORIGINS, which is funded by the Deutsche Forschungsgemeinschaft (DFG, German Research Foundation) under Germany's Excellence Strategy -- EXC-2094 -- 390783311.
MS acknowledges support by the P.~E.~Fil\'en fellowship and a fellowship at the Swedish Collegium for Advanced Study (SCAS).
LM acknowledges support from PRIN MIUR 2017 WSCC32 ``Zooming into dark matter and proto-galaxies with massive lensing clusters''.
AR acknowledges funding from Italian Ministry of Education, University and Research (MIUR) through the ‘Dipartimenti di eccellenza’ project Science of the Universe. He is supported in part by the project “Combining Cosmic Microwave Background and Large Scale Structure data: an Integrated Approach for Addressing Fundamental Questions in Cosmology”, funded by the MIUR Progetti di Rilevante Interesse Nazionale (PRIN) Bando 2017 - grant 2017YJYZAH 
We  acknowledge  use  of  the  Python  libraries \texttt{NumPy} \citep{numpy}, \texttt{Matplotlib} \citep{Matplotlib} and \texttt{ChainConsumer} \citep{ChainConsumer}. This work has made use of Cosmo-Hub \citep{cosmohub2017,cosmohub2020}. CosmoHub has been developed by the Port d’Informació Científica (PIC), maintained through a collaboration ofthe Institut de Física d’Altes Energies (IFAE) and the Centro de Investigaciones Energéticas, Medioambientales y Tecnológicas (CIEMAT) and the Institute of Space Sciences (CSIC \& IEEC), and was partially funded by the ``Plan Estatalde Investigación Científica y Técnica y de Innovación'' program of the Spanish government.
\AckEC
\end{acknowledgements}

\bibliographystyle{aa}
\bibliography{biblio}

\clearpage
\newpage

\begin{appendix}

\section{Void count measures}\label{AppendixA}

\Cref{tab:void_counts_equispaced} provides the number of voids identified in the redshift-space distribution of galaxies, in different redshift bins. We note that, conversely to \Cref{tab:void_counts}, it relies on generic equi-spaced $\Delta z = 0.1$ bins, to facilitate the use for future forecasts of different void applications and combination with other probes. Here we report void number counts obtained both before and after the application of the cleaning procedure. Moreover, we show the number of cosmic voids considering both an optimistic and a pessimistic cut on smaller void radii, that is voids with radius over $1$ time the MGS, and voids over $2$ times the MGS, respectively. The lowering of void counts in the outermost bins is caused by survey mask effects at redshift boundaries of the simulated light-cone.

\begin{table*}
\caption{Void counts in nine equi-spaced bins in redshift, measured in the redshift-space mock galaxy catalogue, provided as a reference for future forecast analyses. The first column represents the minimum and the maximum redshift values for each bin, while the second and the third columns provide the volume in units of $(h^{-1} \ \mathrm{Gpc})^3$ corresponding to each shell of the sky octant, and the MGS, respectively. The next two columns show the number of voids identified by the \texttt{VIDE} void finder, selected with an effective radius greater than $1$ and $2$ times the MGS, respectively. In the last two columns we provide the void number counts obtained after the cleaning procedure, to be modelled using the void size function theory. The latter are reported with the same radius selections as described before. In the last row we show the total shells' volume, the mean MGS and the total void counts corresponding to the entire range of redshifts.}

\centering
\begin{tabular}{cccccccccc}
\toprule
\multirow{2}{*}{$z$ range} & \multirow{2}{*}{shell volume $[(h^{-1} \ \mathrm{Gpc})^3]$} & \multirow{2}{*}{MGS $[h^{-1} \ \mathrm{Mpc}]$} & &
\multicolumn{2}{c}{all voids} & & \multicolumn{2}{c}{voids after cleaning} \\
\cline{5-6} \cline{8-9} \noalign{\smallskip}
& & & & $R>\mathrm{MGS}$ & $>2 \, \mathrm{MGS}$& & $R_\text{eff}>\mathrm{MGS}$ & $>2 \, \mathrm{MGS}$ \\
\midrule
$0.9-1.0$  & $1.308$ & $10.28$ & & $8928$ & $6032$ & & $4845$  & $726$   \\
$1.0-1.1$  & $1.427$ & $11.02$ & & $8987$ & $6637$ & & $5253$  & $840$   \\
$1.1-1.2$  & $1.531$ & $11.74$ & & $7735$ & $5824$ & & $4690$  & $699$   \\
$1.2-1.3$  & $1.622$ & $12.63$ & & $7167$ & $5237$ & & $4140$  & $500$  \\
$1.3-1.4$  & $1.700$ & $13.51$ & & $6575$ & $4756$ & & $3703$  & $321$   \\
$1.4-1.5$  & $1.766$ & $14.45$ & & $5636$ & $4078$ & & $3152$  & $249$   \\
$1.5-1.6$  & $1.821$ & $15.45$ & & $5132$ & $3624$ & & $2719$  & $160$   \\
$1.6-1.7$  & $1.867$ & $16.48$ & & $4389$ & $3049$ & & $2286$  & $94$   \\
$1.7-1.8$  & $1.904$ & $17.63$ & & $2248$ & $934$ & & $851$  & $4$    \\
\midrule
$0.9-1.8$  & $14.95$ & $13.69$ & & $56\,797$ & $40\,171$ & & $31\,639$  & $3593$    \\
\noalign{\vspace{0.01cm}}
\hline 
\bottomrule

\end{tabular}
\label{tab:void_counts_equispaced}
\end{table*}

\section{Different setting forecasts}\label{AppendixB}

\begin{figure*}
\centering
\includegraphics[width=0.46\textwidth]{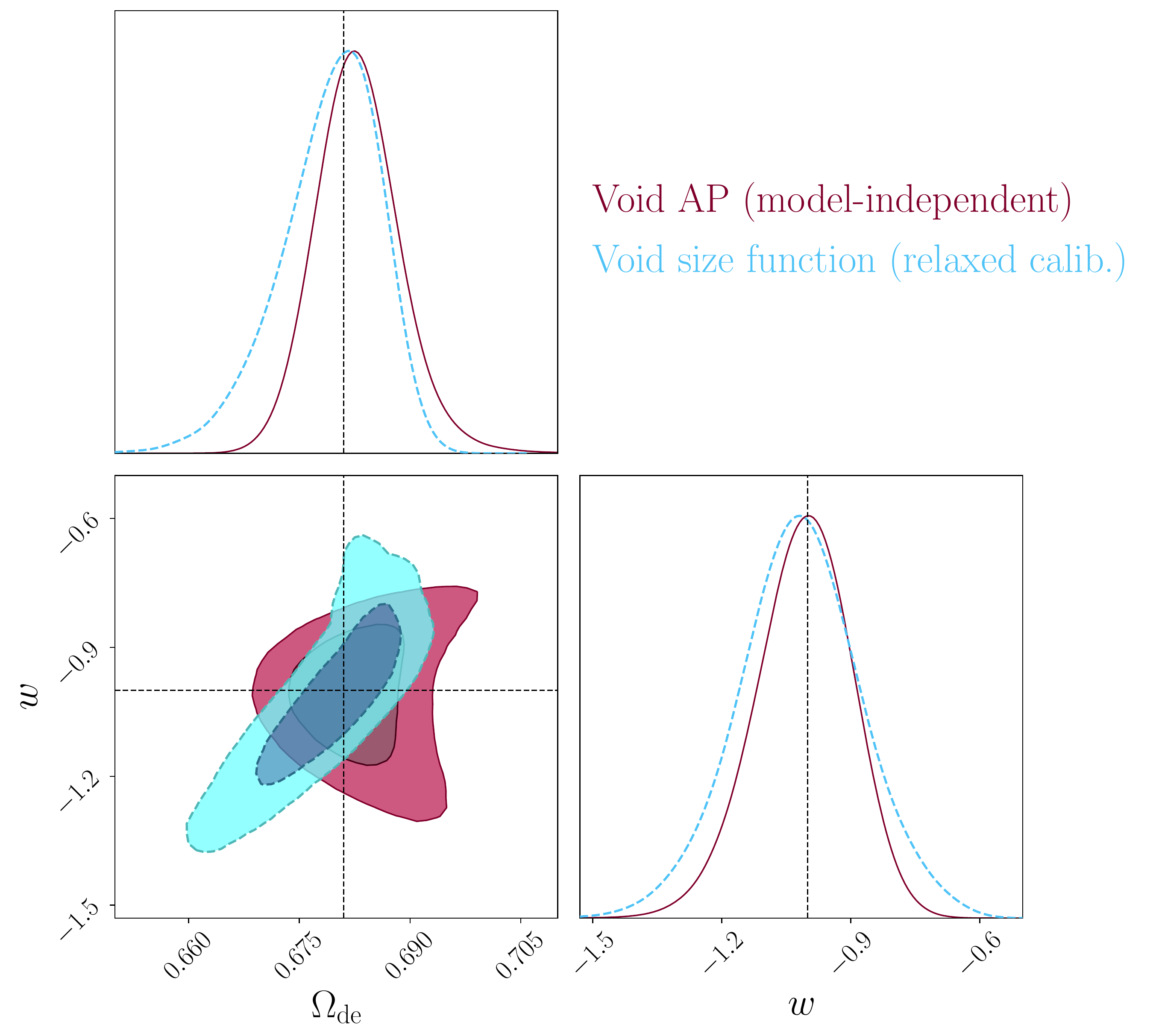}
\hspace{0.25cm}
\includegraphics[width=0.47\textwidth]{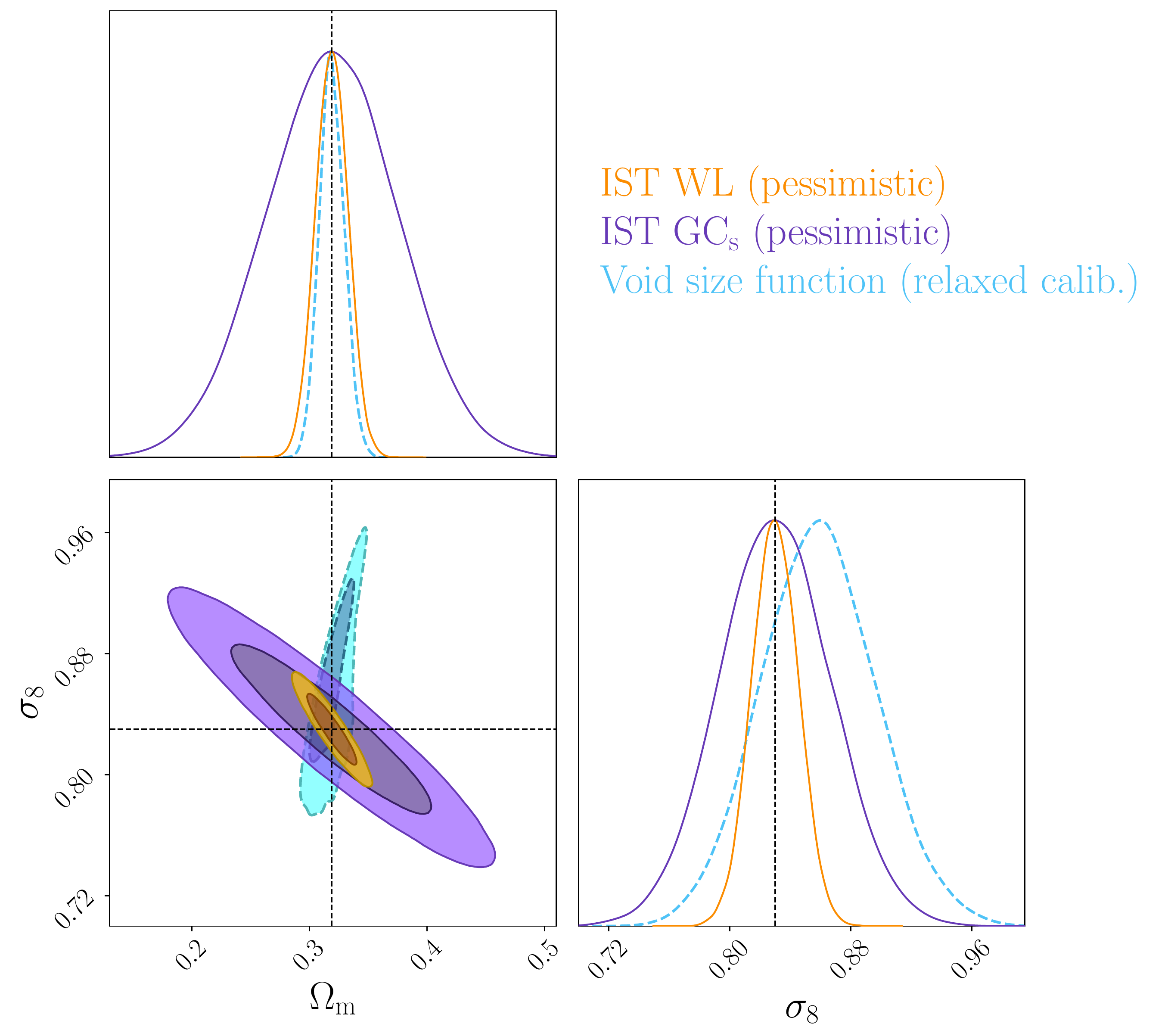}
\caption{Same as \Cref{fig:forecasts_comparison} but for different forecast settings. In this case the confidence contours obtained in this work from the void size function model (light-blue contours with dashed lines) are computed relaxing the constraints given by calibration parameters. The \Euclid forecasts derived with void cross-correlation are computed with a model-independent approach, while IST forecasts are computed with the pessimistic setting described \citep{EC2020}.}
\label{fig:forecasts_comparison_2}
\end{figure*}

We show in \Cref{fig:forecasts_comparison_2} the same forecast comparison presented in \Cref{fig:forecasts_comparison} but using less optimistic settings for the analyses. 
In particular, we report here the \Euclid forecasts on a flat $w$CDM and $w_0w_a$CDM cosmology with massless neutrinos, in the left and right panels respectively, indicating the void size function constraints as light-blue contours with dashed borders.

In the left panel of \Cref{fig:forecasts_comparison_2} we compare our relaxed-calibration results (see \Cref{subsec:cosmological_models} and \Cref{subsec:Vdn_analysis}), with the \Euclid model-independent forecasts of \cite{EuclidHamaus2021}, represented in magenta with solid borders. The latter are computed by means of the void-galaxy cross-correlation function in redshift space, with a calibration-independent approach: the two nuisance parameters of the model are left free to vary and therefore are constrained by the data directly (instead of being fixed to the values calibrated with mock catalogues, i.e. the Flagship light-cone). The authors emphasize that the calibration-independent approach is to be preferred, as fixing the nuisance parameters to the mock values may introduce a prior dependence on the model parameters assumed in the mocks, possibly yielding biased cosmological constraints and underestimated relative uncertainties.

Finally, in the right panel of \Cref{fig:forecasts_comparison_2} we show the comparison of our relaxed-calibration constraints with the Fisher forecasts computed with the pessimistic setting described in \citet{EC2020} for the probes weak lensing ($\mathrm{WL}$, in orange) and spectroscopic galaxy clustering ($\mathrm{GC_s}$, in purple). According to the authors, the pessimistic configuration used for these constraints differs from the optimistic one mainly for a stronger cut of the maximum angular mode for the weak lensing angular power spectrum, $\ell_{\rm max}(\mathrm{WL})=1500$ ($5000$ in the optimistic setting), and of the power spectrum maximum scale, $k_{\rm max}(\mathrm{GC_s}) = 0.25 \ h^{-1} \ \mathrm{Mpc}^{-1}$ ($ 0.3 \ h^{-1} \ \mathrm{Mpc}^{-1}$ in the optimistic setting).
Analogously to \Cref{fig:forecasts_comparison}, the presented confidence contours marginalised on the analysed parameter space are comparable and partially complementary.

\end{appendix}

\end{document}